\def\draftversion{false}
\RequirePackage{ifthen}
\ifthenelse{\equal{\draftversion}{true}}{
  \documentclass[aps,prx,10pt,galley,amsmath,amssymb,showpacs,
                 superscriptaddress]{revtex4}
}{
  \documentclass[aps,prx,10pt,twocolumn,showpacs,
                 superscriptaddress]{revtex4-1}
}

\usepackage{times}
\usepackage{amsmath}
\usepackage{amssymb}
\usepackage{graphicx}
\usepackage[usenames, dvipsnames]{color} % colors
\usepackage{bm} % bold math
\usepackage{subfigure}

\usepackage{ulem}  % for strike-out \sout
\newcommand{\angstrom}{\mbox{\normalfont\AA}}

\def\I{\uppercase\expandafter{\romannumeral 1}}
\def\II{\uppercase\expandafter{\romannumeral 2}}
\def\III{{\uppercase\expandafter{\romannumeral 3}}}
\def\IV{{\uppercase\expandafter{\romannumeral 4}}}
\def\V{{\uppercase\expandafter{\romannumeral 5}}}
\def\k{\mathbf{k}}

\def\nn{\nonumber\\}
\def\qp{\mathbf{q}_{\parallel}}
\def\kp{\mathbf{k}_{\parallel}}
\def\kps{k_{\parallel}}
\def\qps{q_{\parallel}}
\def\tt{\widetilde{t}}
\def\tw{\widetilde{\omega}}

\def\tmu{\widetilde{\mu}}
\def\tnu{\widetilde{\nu}}
\def\RPA{\textrm{RPA}}
\def\Im{\textrm{Im}}
\def\Green#1{\textcolor{OliveGreen}{#1}}

\begin{document}

\title{Correlation and transport phenomena in topological nodal-loop semimetals}
\date{today}

\author{Jianpeng Liu}
\affiliation{ Kavli Institute for Theoretical Physics, University of California, Santa Barbara
CA 93106, USA}

\author{Leon Balents}
\affiliation{ Kavli Institute for Theoretical Physics, University of California, Santa Barbara
CA 93106, USA}

\date{\today}

\begin{abstract}
  We study the unique physical properties of topological nodal-loop
  semimetals protected by the coexistence of time-reversal and
  inversion symmetries with negligible spin-orbit coupling.  We argue
  that strong correlation effects occur at the surface of such systems
  for relatively small Hubbard interaction $U$, due to the narrow
  bandwidth of the ``drumhead'' surface states.  In the Hartree-Fock
  approximation, at small $U$ we obtain a surface ferromagnetic phase
  through a continuous quantum phase transition characterized by the
  surface-mode divergence of the spin susceptibility, while the bulk
  states remain very robust against local interactions and remain
  non-ordered. At slightly increased interaction strength, the system
  quickly changes from a surface ferromagnetic phase to a surface
  charge-ordered phase through a first-order transition.  When
  Rashba-type spin-orbit coupling is applied to the surface states, a
  canted ferromagnetic phase occurs at the surface for intermediate
  values of $U$. The quantum critical behavior of the surface
  ferromagnetic transition is nontrivial in the sense that the surface
  spin order parameter couple to Fermi-surface excitations from both
  surface and bulk states.  This leads to unconventional Landau
  damping and consequently a na\"ive dynamical critical exponent
  $z\!\approx\!1$ when the Fermi level is close to the bulk nodal
  energy. We also show that, already without interactions, quantum
  oscillations arise due to bulk states, despite the absence of a
  Fermi surface when the chemical potential is tuned to the energy of
  the nodal loop. The bulk magnetic susceptibility diverges
  logarithmically whenever the nodal loop exactly overlaps with a
  quantized magnetic orbit in the bulk Brillouin zone.  These
  correlation and transport phenomena are unique signatures of nodal
  loop states.
\end{abstract}


\maketitle

%-------- MARGIN COMMENTS --------------
\def\scr{\scriptsize}
\ifthenelse{\equal{\draftversion}{true}}{
  \marginparwidth 2.7in
  \marginparsep 0.5in
  \newcounter{comm} % counter for commentaries
  % increase counter
  \def\commnext{\stepcounter{comm}}
  % commentary in text
  \def\commtext{{\bf\color{blue}[\arabic{comm}]}}
  % commentary in margin
  \def\commmar{{\bf\color{blue}[\arabic{comm}]}}
  % comment commands for all authors
  \def\lbm#1{\commnext\marginpar{\small LB\commmar: #1}\commtext}
  \def\jlm#1{\commnext\marginpar{\small JPL\commmar: #1}\commtext}
  \def\mlab#1{\marginpar{\small\bf #1}}
  \def\tnewpage{\newpage\marginpar{\small Temporary newpage}}
  \def\tfootnote#1{\Green{\scr [FOOTNOTE: #1]}}
}{
  \def\lbm#1{}
  \def\jlm#1{}
  \def\mlab#1{}
  \def\tnewpage{}
  \def\tfootnote#1{\footnote{#1}}
}
%----------------------------------

%%%% Introduction, review experiments, material proposals and give an overview.

The theoretical proposal and experimental verification of Weyl and Dirac semimetals
\cite{nielsen-plb83,burkov-prl11,wan-prb11,murakami-prb08,halasz-prb12,
turner-13-review,hosur-13-review,jpliu-prb14-a,taas-theory-dai,taas-exp-ding1,taas-exp-ding2,
taas-exp-hasan,typeii-wsm,mote2-exp-liang16,na3bi-theory,na3bi-exp,cd3as2-theory,cd3as2-exp} 
has shown that topological electronic structure is not restricted to gapped systems \cite{kane-rmp10,zhang-rmp11,snte-tci-theory,fu-tsc-prl08,qi-tsc-prl09}, 
but also occurs in gapless systems 
such as nodal metals\cite{burkov-prb11}. 
Recently, the interest in topological semimetals has been extended
from systems with point nodes to those with
a 3D nodal loop,  ``nodal-chain" \cite{nodal-chain-arxiv16},
``nodal-arc"\cite{ptsn4-exp}, and even ``nodal surfaces" \cite{weng-prb16}, 
in which there are bulk band touchings along
isolated or connected 1D lines, or even at 2D surfaces in the 3D Brillouin zone (BZ) 
instead of at isolated points.  

A growing number of material systems  have been theoretically proposed to realize nodal-loop 
semimetals (NLSMs) \cite{graphite-nl-volovik13, nl-mtc-prb15, cu3pdn-theory-prl15,
zrsis-exp, ca3p2-theory, pbtase2-exp, blackp-nl-theory, nl-murakami16, huang-prb16}. In particular, 
ZrSiS and PbTaSe$_2$ have been experimentally confirmed
by angle-resolved photoemission spectroscopy (ARPES) measurements, and the 
bulk nodal loops in the ZrSiS-family compounds were further investigated by 
de Has-van Alphen (dHvA) quantum oscillations \cite{zrsis-qo1,zrsis-qo2} 
and magneto-transport measurements \cite{zrsis-mr}.

In this paper, we discuss some fundamental physics of NLSMs which is
distinct from Weyl and Dirac systems.  First, we argue that nodal-loop
semimetals are prime candidates to observe correlation effects at
their surfaces.  This is because, unlike point node materials which
possess highly dispersive bulk and surface states (typically with
large Fermi velocities derived naturally from the several eV width of
the associated bands), nodal-loop semimetals possess ``drumhead"-like
surface states. Depending on surface terminations, the states exist
either inside or outside the projection of the nodal loop in the
surface BZ.  

The dispersion of such drumhead surface states is
typically much smaller than that of the bulk valence and conduction
bands, raising the interesting possibility of correlation effects
occurring at the surface even when interactions are too weak to
disturb the electronic states with large kinetic energy in the
interior of the sample.  Correlations may be induced by Coulomb
interactions and/or coupling to phonons, due to the small kinetic
energy and large surface density of states.  For example, it has been
theoretically proposed that such novel flat surface states might
support $s$-wave superconductivity whose critical temperature scales
linearly with the coupling strength
\cite{volovik-surfsc1,volovik-surfsc2,surfsc-graphite}.  Here we argue
that repulsive Coulomb interactions generate unusual surface charge
density wave and ferromagnetic states, for moderate interaction
strength for which the bulk states are unaffected.  We expound this in
detail through a thorough Hartree-Fock study of a NLSM, including both
Hubbard $U$ and surface Rashba-like spin-orbit coupling (SOC)\cite{rashba84} .  This
yields a phase diagram showing several correlated surface phases at relatively
small values of $U$.   

Given the prospect for surface quantum phase transitions (QPTs) in
these systems, it is interesting to explore the associated quantum
critical behavior.  We find that such {\em surface} QPTs can realize
entirely new critical universality classes different from either two
or three-dimensional bulk QPTs, owing to their mixed
dimensional character.  Specifically, a distinct process of Landau
damping of order parameter fluctuations into
the third dimension arises,  and dominates under conditions which we
explain.  

It is also important to be able to characterize a NLSM by probes other
than photoemission, which may be difficult or impossible on many
samples, or on appropriate crystal surfaces.  In that vein, we derive
the existence of unconventional quantum oscillations in NLSMs, which
are present even when the Fermi level is exactly at the degeneracy
level, so that the system has no true Fermi surface.

These results are expounded in detail in the remainder of the paper,
which is organized as follows.  In Sec.~\ref{sec:tbmodel}, we first a
noninteracting tight-binding (TB) model on a tetragonal lattice with
both inversion ($\mathcal{P}$) and time-reversal ($\mathcal{T}$)
symmetries, which can realize the NLSM phase when spin-orbit coupling
(SOC) is neglected.  Then, in Sec.~\ref{sec:hubbard} we apply on-site
Hubbard interactions (the strength of the interaction is denoted by
$U$), and solve such an interacting model in a slab geometry within
the Hartree-Fock (HF) approximation, both with and without Rashba SOC,
and complement the HF analysis with a study of the susceptibility in
the random-phase approximation.  Next, in Sec.~\ref{sec:qc}, we
consider Landau damping of ferromagnetic surface fluctuations, which
control quantum critical phenomena \cite{hertz-prb76, millis-prb93}.
We find in particular that when the Fermi level is close to the nodal
energy, the dominant process is one in which an electron-hole pair is
{\em shared} between the bulk and surface, leading to an
unconventional dynamical coefficient
$\sim\!\vert\nu_m\vert q_{\parallel}$ ($\nu_m$ is the bosonic
Matsubara frequency, $\qps$ is the magnitude of in-plane wavevector).
This implies a new universality class for the ferromagnetic QPT.
Finally, In Sec.~\ref{sec:qo}, we discuss quantum oscillations due to
the bulk nodal-loop states, showing that they arise even in the
absence of a Fermi surface, and conclude with a summary in
Sec.~\ref{sec:summ}.

%%-----------------------------------------------------
\section{Non-interacting tight-binding model}
\label{sec:tbmodel}
%%------------------------------------------------------

\begin{figure}
\includegraphics[width=3.0in]{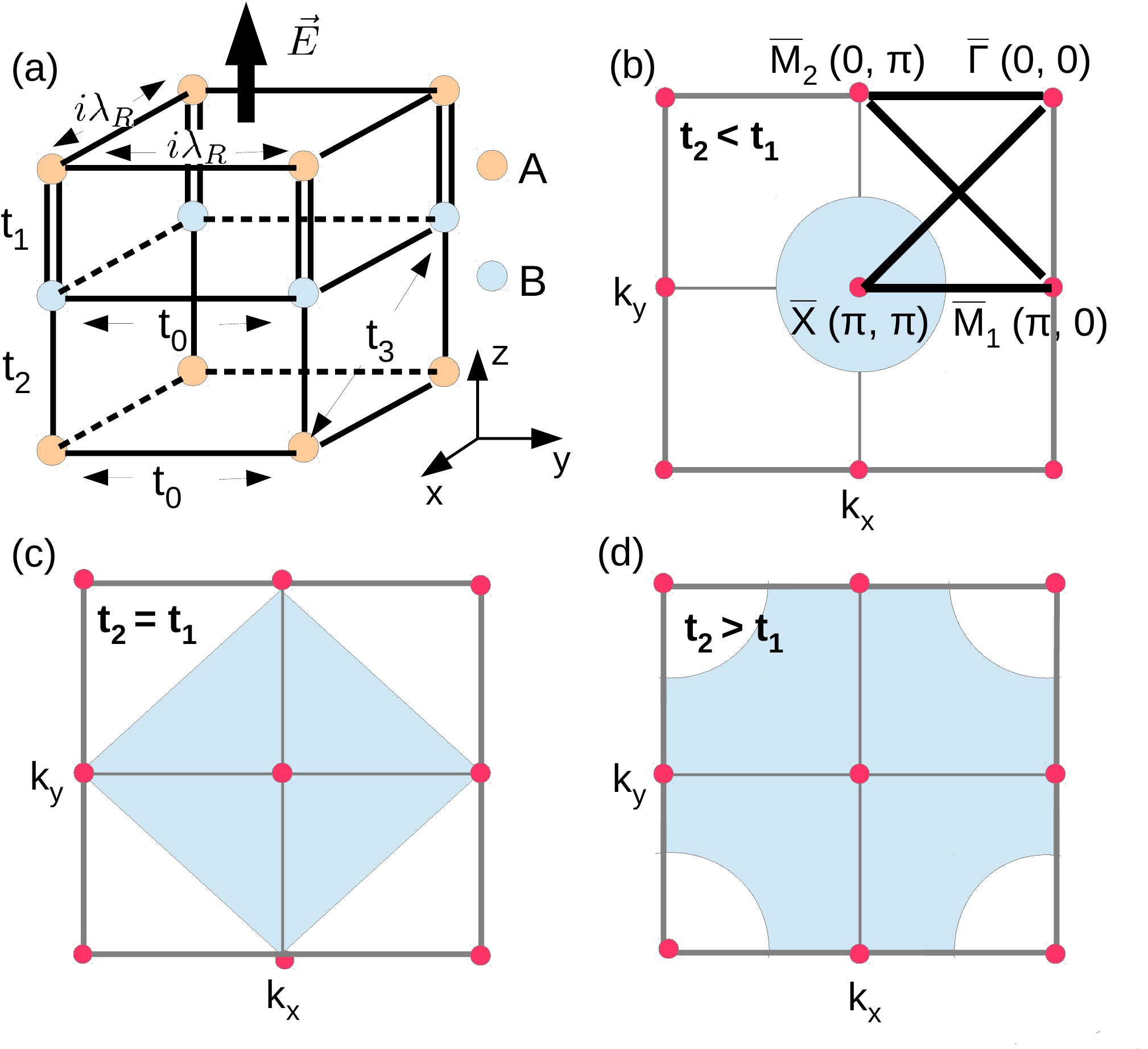}
\caption{Schematic illustration of the non-interacting tight-binding model
for nodal loop semimetals on a tetragonal lattce. (a) Lattice structure
and hoping terms, the thick black arrow indicate surface electric field
which generates Rashba SOC denoted by $\lambda_R$ (b)-(d), nodal loops 
projected onto the (001) surface BZ, with the shaded region indicating
the drumhead surface states, (b) for $t_2\!<\!t_1$, (c) $t_2\!=\!t_1$, and
(d) $t_2\!>\!t_1$}
\label{fig:tb}
\end{figure}

We first construct a non-interacting TB model on a tetragonal lattice with both $\mathcal{T}$ and
$\mathcal{P}$ symmetries neglecting SOC. As schematically 
shown in Fig.~\ref{fig:tb}(a),  there are two sublattices
denoted by $A$ and $B$ in each primitive cell, and the hopping from $A$ to $B$ along the positive (negative)
$z$ direction is denoted by $t_1$ ($t_2$). Moreover, there are intra-sublattice in-plane hopping
$t_0$ and inter-sublattice in-plane hopping $t_3$.  
Without the in-plane hoppings, the system can be
considered as arrays of decoupled 1D Su-Schrieffer-Heeger (SSH) chains \cite{ssh-1,ssh-2};
the in-plane hoping $t_3$ couple these
chains together so that there is band inversion
around only one of the eight time-reversal invariant momenta (TRIM).
The nodal loop is centered around the TRIM with inverted band order.  

\begin{figure}[th]
\includegraphics[width=3.2in]{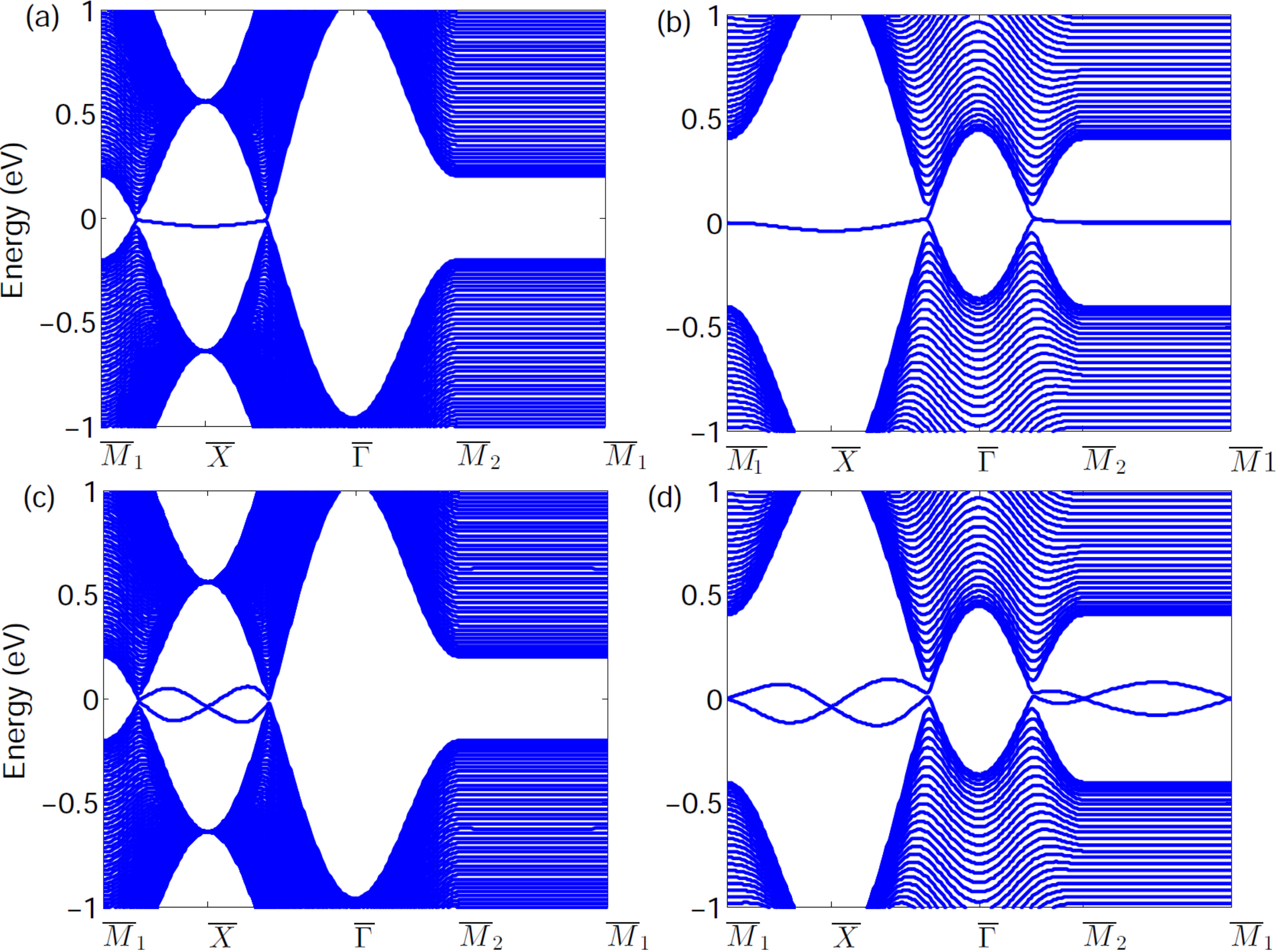}
\caption{Surface bandstructures of the non-interacting tight-binding model without surface SOC (a)-(b),
and with surface SOC (c)-(d). (a) $t_2\!=\!0.75t_1$, and (b) $t_2\!=\!1.25t_1$; (c) $t_2\!=\!0.75t_1$,
$\lambda_R\!=\!0.0625t_1$, and (d) $t_2\!=\!1.25t_1$,
$\lambda_R\!=\!0.0625t_1$. The energy bands are plotted along the high-symmetry path marked by
the thick black lines in Fig.~\ref{fig:tb}(b).}
\label{fig:surfband}
\end{figure}

The specific properties of  the nodal loop such as its size and shape are controlled 
by $t_1$, $t_2$ and $t_3$, 
while $t_0$ renders dispersions to both the bulk nodal energy along the loop and
the otherwise flat drumhead surface states.
Hereafter we fix $t_1\!=\!0.8$, $t_3\!=\!0.2$, $t_0\!=\!0.01$, and $t_2\!>\!0$ is 
the only variable in the noninteracting situation.  In particular, when $t_2\!<\!t_1$, there is
a circular nodal loop centered at the $X$ ($(\pi, \pi,\pi)$) point. If the surface is truncated
at the $A$ sublattice, one obtains drumhead surface states inside the projected nodal
loop centered at $\overline{X}$ as shown in Fig.~\ref{fig:tb}(b) and Fig.~\ref{fig:surfband}(a). 
If  $t_2\!=\!t_1$,  the nodal loop is diamond-like and connects
the TRIM $X$ and $M$ ($(\pi,0,\pi)$). The corresponding 
surface states fill the region inside the diamond as
shown in Fig.~\ref{fig:tb}(c)
\footnote{Even though the shape of the nodal loop looks perfectly nested when $t_1\!=\!t_2$, 
the Fermi surface is not nested due to the dispersion of the nodal energy from $t_0$.}.
When $t_2\!>\!t_1$, the nodal loop is centered
at $Z$ ($(0,0,\pi)$) and the surface states fill the region outside the projected
nodal loop (Fig.~\ref{fig:tb}(d) and Fig.~\ref{fig:surfband}(b)). It worth to note that for fixed bulk hopping parameters 
the drumhead surface states can be either inside or outside the projected nodal
loop depending on surface terminations (see Appendix \ref{appen:gsf}), which is essentially due to the properties of 1D SSH chains. 
Therefore, the surface states covering a large portion of the surface BZ as 
shown in Fig.~\ref{fig:tb}(d) can also be realized when $t_1\!<\!t_2$ 
if the system is terminated at the other sublattice.

Given that inversion symmetry is always broken at a surface, the surface electric field
may lead to considerable Rashba spin-orbit splittings in the surface states. Such surface Rashba splittings
have been observed in the surfaces of 
nonmagnetic and magnetic metals \cite{Au111_prl96,Bisurface_prl04,rashba-magnetimetal-prb05}, as well as 
semiconductor heterostructures \cite{soc-splitting-enoki-prl97}. Thus we also take the surface
Rashba effects into account by adding a Rashba-type first-neighbor spin-dependent hopping
within the surface atomic layer,  of which the amplitude is denoted by $\lambda_R$. 
The spin-degenerate drumhead surface states are splitted
by such surface SOC (see Fig.~\ref{fig:surfband}(c)-(d)); moreover, 
the surface states acquire nontrivial spin textures.
We thus expect that the effects of Coulomb interactions in 
these two situations (with and without surface SOC) would be 
different.

%%---------------------------------------------------------------
\section{Effects of Hubbard interactions}
\label{sec:hubbard}
%%---------------------------------------------------------------

%%----------------------------------------------------------------
\subsection{Without surface Rashba spin-orbit coupling}
\label{sec:huubardnosoc}
%%----------------------------------------------------------------

We first consider the situation without surface Rashba splittings, and apply Hubbard interactions, $H_{U}\!=\!U\sum_{i}\hat{n}_{i\uparrow}\hat{n}_{i\downarrow}$, 
to the above noninteracting tight-binding model in a slab geometry. 
As the Coulomb interaction at the surface
is expected to be strongly screened due to the large surface density of states (DOS), 
a Hubbard-type local interaction is a good description if we are mainly 
interested in the effects on the surface 
states. On the other hand, unlike the surface states of topological insulators, 
there is no simple low-energy effective Hamiltonian
describing the drumhead surface states of NLSMs. Thus we have to construct a slab and apply
Hubbard interactions to all the electrons in the slab.
Hereafter we will only consider half-filled systems, 
and we say the system is charge homogeneous with zero charge density 
if each site is exactly half filled, i.e., there is one electron at each site.

\begin{figure}
\includegraphics[width=2.5in]{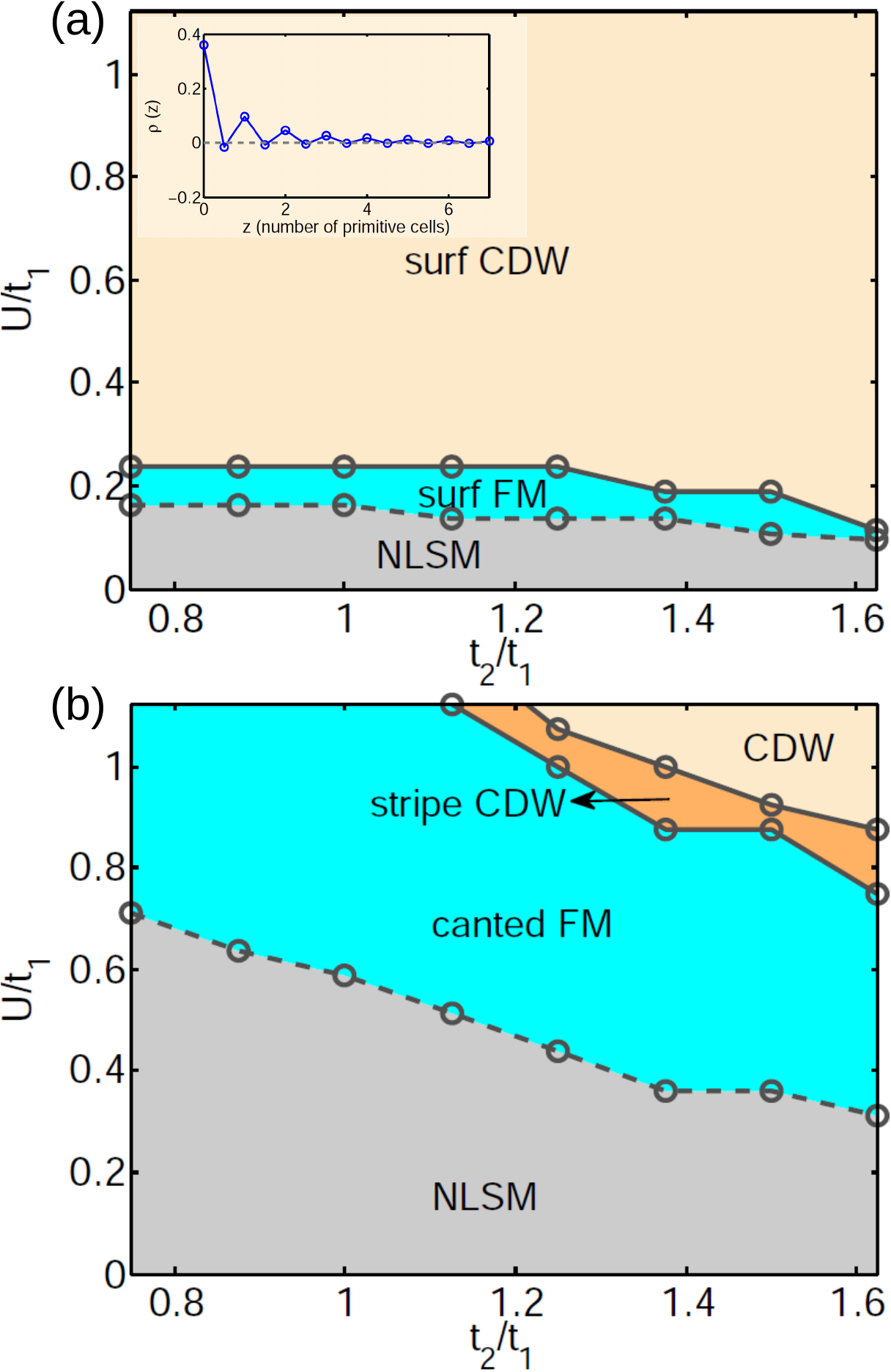}
\caption{ Phase digram of the NLSMs with Hubbard interactions in the $t_2-U$ parameter space:
(a)Without surface Rashba SOC, with the inset shows the local charge density distribution in the surface CDW phase 
when $t_2\!=\!1.25t_1$ and $U\!=\!0.5t_1$; and (b) With surface Rashba SOC.}

\label{fig:phasediagram}
\end{figure}

The Hubbard interactions are treated by self-consistent Hartree-Fock (HF) approximation 
(see Appendix (\ref{appen:hf}) for details). 
The HF ground states for a slab of 50 primitive cells 
are shown in Fig.~\ref{fig:phasediagram}(a). When $U\!=\!0$, the system
is in the NLSM phase. When $U\!\sim\!10\%\!-\!20\%\,t_1$, 
the system enters into a surface FM 
(denoted by ``surf FM" in the figure) phase with the 
ferromagnetic order localized at the surface.
As $U$ is further increased, a surface charge-ordered phase becomes energetically favored over the 
surface FM phase. The system enters enter in to surface CDW phase trough a first-order transition.
The inset in Fig.~\ref{fig:phasediagram}(a) shows the local charge 
density along the $z$ direction for $U\!=\!0.5t_1$ 
and $t_2\!=\!1.25t_1$. Clearly the charges are strongly localized at the 
surface, as the density oscillation decays rapidly into the bulk.

To study the nature of the surface FM transition, 
we have calculated the spin susceptibility of a 30-unitcell slab in the 
random phase approximation (RPA) \cite{rpa-spinel} (see Appendix \ref{appen:rpa} for details). 
Fig.~\ref{fig:chi}(a) shows the eigenvalues of static RPA spin susceptibility at different wavevectors 
at $U\!=\!0.25t_1$ and $t_2\!=\!t_1$. 
As clearly shown in the figure, there are a large number of quasi-degenerate bands with small amplitudes; 
moreover, there are two degenerate
bands with much larger amplitudes which tend to diverge at $\Gamma$.  
The eigenvectors of the RPA spin susceptibility indicate that 
those quasi-degenerate bands with small amplitudes are from the bulk spin fluctuations, while the
two  bands with much larger amplitudes are dominated by acoustic and optical surface fluctuation modes. 
This is consistent with the expectation that the drumhead surface states are much more 
sensitive to Coulomb interactions than the bulk states due to the much smaller bandwidth. 
From Fig.~\ref{fig:chi}(a) it is also evident that the surface spin-fluctuation modes tend to 
diverge at $\overline{\Gamma}\!=\!(0,0)$, indicating a continuous quantum phase transition at the 
surface driven by Hubbard interactions. We refer the readers to Appendix \ref{appen:rpa}
for technical details of the implementation of RPA on the slab 
as well as the properties of the eigenvalues and the eigenvectors 
of the spin susceptibility.

%Since the surface  charge fluctuations of $\chi^{\rho}(\q)$ are much stronger than the bulk  
%modes,  it deserves further discussions.
In Fig.~\ref{fig:chi}(b) we show  
the parameter dependence of the RPA surface spin susceptibility at $\overline{\Gamma}\!=\!(0,0)$ 
(denoted by $\chi_{zz}^{\text{surf}}(\Gamma)$). As is clearly seen from the figure, for a given $t_2$,  
the surface fluctuation modes at $\overline{\Gamma}$ increase with $U$, and diverge
at some critical $U$, indicating the transition from a 
nonordered phase to a surface FM phase. The gray dotted line in Fig.~\ref{fig:phasediagram}(b) marks
the numeric threshold above which $\chi_{zz}^{\text{surf}}(\Gamma)$ is considered as diverging.
It is interesting to note that 
as $t_2$ increases from $0.75t_1$ (denoted by blue crosses)  to $1.5t_1$ (denoted by cyan diamonds), the 
critical $U$ value is reduced by $\sim$50\%. 
This is because the surface DOS becomes larger for greater $t_2$ values (Fig.~\ref{fig:tb}(b)-(d)),
thus the system becomes more sensitive to Coulomb interactions. 

\begin{figure}
\includegraphics[width=2.5in]{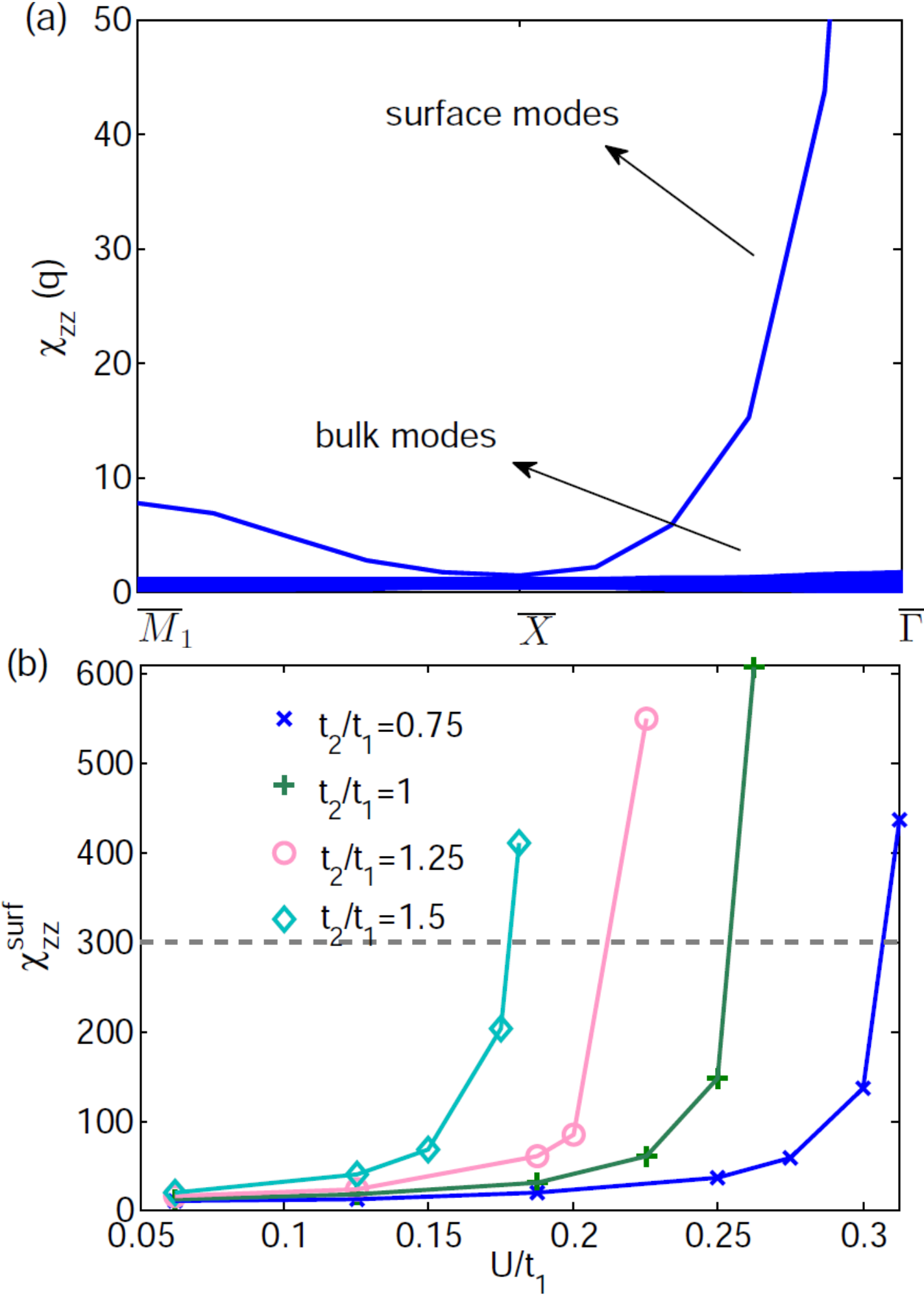}
\caption{(a) Dispersion of the spin susceptibility ($\chi_{zz}(q)$) 
for a 60-layer slab of nodal-loop metal with $t_2=t_1$ and $U=0.25 t_1$. 
(b) The $U$ dependence of the surface spin fluctuations at $\Gamma$ 
(denoted by $\chi^{surf}_{zz}$)  for different $t_2$ values.}
\label{fig:chi}
\end{figure}
%%

%%---------------------------------------------------------
\subsection{Hubbard interactions with surface Rashba SOC}
\label{sec:hubbardsoc}
%%----------------------------------------------------------

We continue to study the effects of Hubbard interactions 
on NLSMs including surface Rashba splittings with $\lambda_{\textrm{R}}\!=\!0.0625t_1$.
Since the surface electric field decays quickly into the bulk, 
it is assumed that the Rashba SOC $\lambda_\textrm{R}$ applies 
only to the topmost and bottommost layers of the slab.  
The system with such surface SOC expects to be more robust
against Coulomb interactions due to the 
lifted spin degeneracy of the drumhead surface states as shown in Fig.~\ref{fig:surfband}(c)-(d).
Moreover, as the surface states at the Fermi level 
acquire nontrivial spin textures due to Rashba SOC, 
it is unlikely that a charge-ordered phase would be favored.

Both of the above two conjectures are numerically verified as shown in Fig.~\ref{fig:phasediagram}(b).
When surface SOC is turned on,
our noncollinear self-consistent HF calculations (see Appendix \ref{appen:hf} for technical details) 
suggest that the system tends to enter into a surface canted FM phase 
around some moderate $U$ values ($U_{\textrm{c}}\!\sim\!35\%\!-\!65\%\,t_1$).
The surface canted FM phase is characterized by ferromagnetically 
coupled $z$ components of spins ($m_z$) which are exponentially localized
at the surface, and possibly with small spin cantings toward the in-plane directions.

We have also checked the $U$ dependence of $m_z$ at the surface layer,
and find that  $\vert m_z\vert$ increases continuously with $U$ when
$U\!\ge\!U_c$, indicating a continuous quantum phase transition. The critical value $U_c$ decreases
with the increase of $t_2$ due to the larger surface DOS for greater $t_2$ values. The continuous 
quantum phase transition is further 
verified by the divergence of surface spin susceptibility (data not shown).
Moreover, it turns out that $\vert m_z\vert$ is likely to have a square root dependence 
on $U-U_c$ ($\vert m_z\vert\!\sim\!\sqrt{U-U_c}$ ), which is in agreement with 
the behavior of Stoner ferromagnetism. 
\cite{coleman-book}. 

When $t_2\!>\!t_1$ the system tends
to go to a surface stripe charge-ordered phase (indicated by ``stripe CDW" in Fig.~\ref{fig:phasediagram}(a)) 
at large $U$ values, in which there are alternating positive and negative charge
stripes along either the $x$ or the $y$ direction.  There is  a transition from such stripe
CDW phase to a surface CDW with homogeneous in-plane charge density as $U$ further increases.
Both of these transitions (from canted FM to stripe CDW phase, and from stripe CDW to in-plane homogeneous
CDW phase) turn out to be first-order transitions whose phase boundaries are 
marked by solid lines as shown in Fig.~\ref{fig:phasediagram}(b). 
 
%%----------------------------------------------------------
\section{Ferromagnetic quantum criticality at the surface}
\label{sec:qc}
%%-----------------------------------------------------------

\subsection{Framework and general considerations}
\label{sec:fram-gener-cons}

In this section we discuss the quantum critical (QC) behavior near the
ferromagnetic transition at the surface of a nodal-loop semimetal
neglecting effects of surface SOC.  The prototypical description of
the quantum phase transition in an itinerant ferromagnet is that of
Hertz-Millis theory \cite{hertz-prb76,millis-prb93}, in which the
system is described by an effective action for the order parameter in
which the itinerancy of the electrons is reflected by a term
representing Landau damping, due to the coupling with Fermi-surface
fluctuations \cite{moriya-12}. The Landau damping gives rise to a term
quadratic in the order parameter with a dynamical coefficient
$\sim\vert\nu_{m}\vert/q$ in the effective action of the spins.  Based
on this, Hertz derived the dynamical critical exponent $z\!=\!3$ for
FM transitions in 2D and 3D Fermi-liquid systems \cite{hertz-prb76}.
The dynamical critical exponent determines the quantum critical
phenomenology such as the dependence of critical temperatures on $U$,
the specific heat, and the crossover behavior from quantum to
classical regime at finite temperatures \cite{hertz-prb76,
  millis-prb93}.  In two dimensions, there are known flaws in the
purely order parameter description, and much theoretical work has gone
into improving
it \cite{lee2009low,metlitski2010quantum,mross2010controlled}.
Nevertheless, the dynamical scaling $z\approx 3$ is believed to still
be quite a good approximation if not exact.

In NLSMs, we have shown in Sec.~\ref{sec:hubbard} that the FM
transition occurs only at the surface and no order occurs in the bulk,
so that one may na\"ively expect purely two-dimensional FM quantum
criticality with $z \approx 3$.  However, in reality the situation is
more complicated due to the gapless bulk states. The electron-hole
excitations which couple to the surface spin order parameter arise
both from the surface bound states and the extended bulk states, which
have an amplitude at the surface.  Given the critical role of Landau
damping in the theory, we may expect that the quantum critical
behavior would be different for such a surface FM transition with
gapless bulk excitations.

\begin{figure}
\includegraphics[width=3.4in]{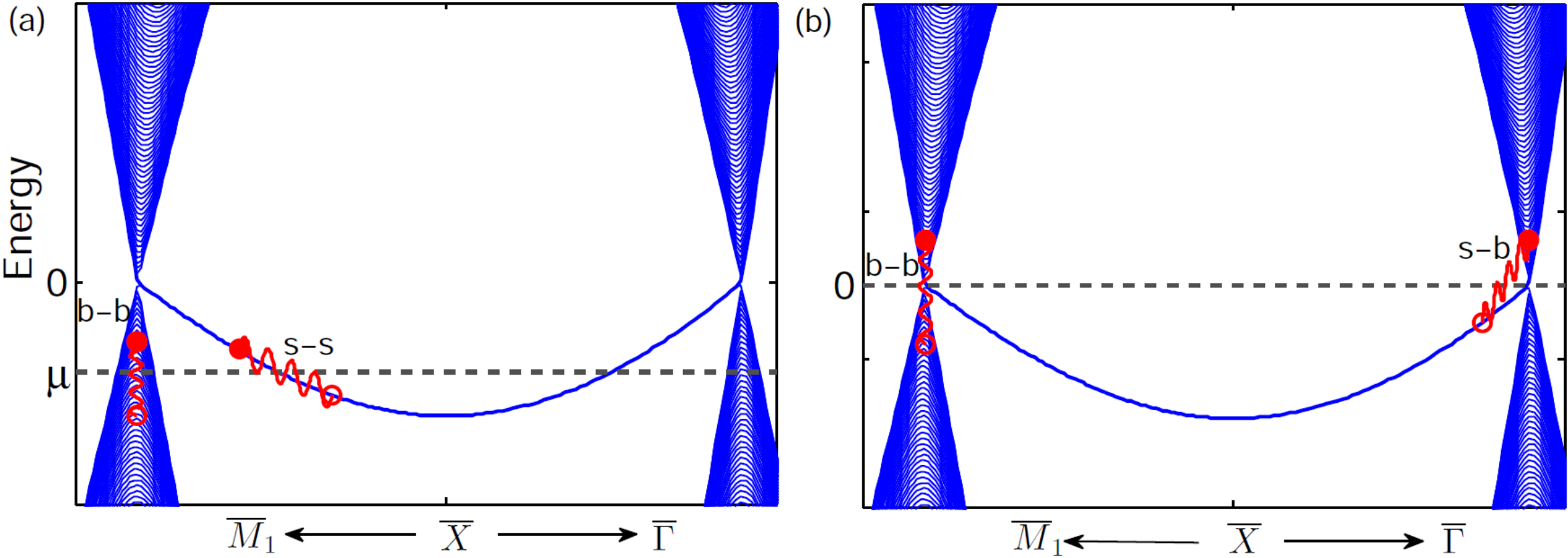}
\caption{Schematic illustration of different types of electron-hole excitations
that couple to surface spins. (a) When the surface bands are partially filled.
(b) When the surface bands are (nearly) completely filled. 
The electron-hole excitations purely from the surface (bulk) states are denoted by ``$s-s$" 
(``$b\!-\!b$"); while the process of creating a hole in the surface states and an electron in the bulk
states is denoted by ``$s\!-\!b$".  }
\label{fig:eh}
\end{figure}

We confine our analysis here to the level of Landau damping, i.e. the
Hertz-Millis order parameter description, which is sufficient to
distinguish the difference between purely 2d critical behavior and
something else.  This is already somewhat subtle because several
distinct processes may contribute to the damping, i.e. the
non-analytic part of the surface spin susceptibility, and one must
carefully take into account the momentum and frequency behavior of
surface Green's functions in describing this.  It is convenient to
decompose the electron-hole excitations into different types. In the
first type, both the electron and the hole are
created in the surface bound states as denoted by $``s\!-\!s"$ in
Fig.~\ref{fig:eh}(a); in the second type, that both the
electron and the hole are created in the bulk continuum which is
denoted as $``b\!-\!b"$ in Fig.~\ref{fig:eh}; and finally in the last
type, a hole is created in the surface states while an
electron is added to the bulk states as denoted by $``s\!-\!b"$ in
Fig.~\ref{fig:eh}(b).

We consider two different situations. The first situation is that the
system is (slightly) hole-doped with partially filled surface bands as
schematically shown in Fig.~\ref{fig:eh}(a).  In the second situation,
the Fermi level is very close to the nodal energy and the drumhead
surface states are almost completely filled as sketched in
Fig.~\ref{fig:eh}(b).  In the first situation we only consider the
$s\!-\!s$ and $b\!-\!b$ type excitations, since the $s\!-\!b$ process
requires a large momentum transfer, and we are only interested in
low-frequency long-wave-length excitations; while in the second case
we only consider the $s\!-\!b$ and $b\!-\!b$ excitations since the
surface bands are fully occupied.

\subsection{Surface Green's function and Dynamical Susceptibility}
\label{sec:surf-greens-funct}

We start by calculating the surface Green's function (SGF) of NLSMs using the 
method reported in Ref.~\onlinecite{kalkstein-71}.   Note that the SGF
includes contributions from both extended and localized
eigenstates, and by using an exact method for calculating the SGF, we
capture subtle behaviors due to varying contributions of the two types
of states.  For the tight-binding model given in Sec.~\ref{sec:tbmodel},
the surface Green's function ($G_{s}(\kp,\omega)$) can be calculated analytically at low energies when the
size of the nodal loop is much smaller than that of the BZ. 
It turns out that the SGF has a simple  analytic solution
\begin{align}
G_s(\kp, \omega)&=\frac{-\tw}{\tt_1(\,(\sqrt{\gamma^2-4}+\gamma)/2-t_2/\tt_1\,)}\;\nn
&\approx-\frac{1}{t_2}\frac{\tw}{\sqrt{(\kps^2-k_0^2)^2-\tw^2}+\kps^2-k_0^2}
\label{eq:sgf}
\end{align}
where $\gamma\!=\!(\tt_1^2+t_2^2-t_2^2\tw^2)/(\tt_1 t_2)$, and
\begin{align}
\tw&=(\,\omega-2t_0(\cos{k_x}+\cos{k_y})+\tmu\,)/t_2;,\nn
   &\approx\!(\,\omega-(t_0(k_x^2+k_y^2)-4t_0)+\tmu\,)/t_2\;,
\label{eq:tw}
\end{align}
where $\tmu\!=\!\mu/t_2$ with $\mu$ being the Fermi level, and 
\begin{align}
\widetilde{t}_1&=t_1+2t_3(\cos{k_x} +\cos{k_y}) \; \nn
&\approx\!t_1-4t_2 + t_2(k_x^2+k_y^2)\;. 
\label{eq:tt1}
\end{align}
We consider the situation that the nodal loop is centered at $(\pi,\pi,\pi)$ the radius of which 
is much smaller than the size of the Brillouin zone, and assume 
that $t_2\!=\!t_3$, which is nothing but saying that the bulk Fermi velocity is isotropic. 
Then the second lines in Eq.~(\ref{eq:tw})-(\ref{eq:tt1})
follow by expanding $\cos{k_x}$ and $\cos{k_y}$ around $k_x\!=\!\pi$ and $k_y\!=\!\pi$.
In Eq.~(\ref{eq:sgf}) $k_0$ is introduced as a parameter characterizing the size of the nodal loop:
\begin{align}
\tt_1-t_2&=t_3(\kps^2-k_0^2)\;\nn
         &=t_2(\kps^2-k_0^2)\;.
\label{eq:k0}
\end{align}
Again, we have assumed that $t_2\!=\!t_3$ so that the bulk Fermi velocity is isotropic.
Starting from Eq.~(\ref{eq:sgf}) it is straightforward to show that when 
$-\vert \kps^2-k_0^2\vert\!\leq\!\tw\leq\vert \kps^2-k_0^2\vert$, $\omega$ is the in the bulk gap, and there is 
a pole at $\tw\!=\!0$ for $\kps\!<\!k_0$ corresponding to the drumhead surface states 
(the surface is prepared by making a truncation at the $A$ sublattice) ; 
while when $\tw\!>\!\vert\kps^2-k_0^2\vert$ or $\tw\!<\!\vert\kps^2-k_0^2\vert$, $\omega$ is in the bulk continuum.
Hereafter we will set the bulk nodal energy as 0, so $\tw$ is shifted by a small constant: 
$t_2\tw\!=\!\omega-t_0(\kps^2-k_0^2)+\tmu$. We refer the readers to Appendix \ref{appen:gsf} for details in 
calculating the surface Green's function.

Eq.~(\ref{eq:sgf}) may be expressed using the spectral representation as:
\begin{equation}
G_s(\kp,\omega)=\frac{1}{t_2}\int d\epsilon\,\frac{f(\kp,\epsilon)}{\omega/t_2-(\epsilon-\tmu)+i\delta_{\epsilon}}\;,
\label{eq:gsf-spec}
\end{equation}
where $\delta_{\epsilon}$ is an infinitesimal quantity which is greater than (less than) zero 
if $\epsilon\!>\!\tmu$ ($\epsilon\!<\!\tmu$). Or, in the Matsubara formalism,
\begin{equation}
G_s(\kp,i\omega_n)=\frac{1}{t_2}\int d\epsilon\,\frac{f(\kp,\epsilon)}{i\omega_n/t_2-(\epsilon-\tmu)}\;,
\end{equation}
The spectral density $f(\kp,\epsilon)$ consists of two terms:
\begin{equation}
f(\kp,\epsilon)=f_{b}(\kp,\epsilon)+f_s(\kp,\epsilon)\;.
\label{eq:fsfb1}
\end{equation}
$f_b(\kp,\epsilon)$ is from the bulk continuum, and $f_s(\kp,\epsilon)$ corresponds to the surface bound state:
\begin{align}
&f_{b}(\kp,\epsilon)=\frac{\sqrt{(\epsilon-\tt_0 x_{\kps})^2- x_{\kps}^2}}{\epsilon-\tt_0 x_{\kps}}\theta(\vert\epsilon-\tt_0 x_{\kps}\vert-\vert x_{\kps}\vert)\;,\nn
&f_{s}(\kp,\epsilon)=\vert x_{\kps}\vert\delta(\epsilon-\tt_0 x_{\kps})\theta(-x_{\kps})\;,
\label{eq:fsfb2}
\end{align}
where $x_{\kps}\!=\!\kps^2-k_0^2$, and $\tt_0\!=\!t_0/t_2$. %We introduce three dimensionless parameters which will be used frequently in the remainder of
%the paper:
%
%, and $\tt_0\!=\!t_0/t_2$. 

Now it is straightforward to calculate the dynamical susceptibility using the surface Green's function 
shown in Eq.~(\ref{eq:gsf-spec})-(\ref{eq:fsfb2}). To be specific, using the Matsubara formalism,
the dynamical susceptibility is expressed as:
\begin{equation}
\chi(\qp,i\nu_m)\!=\!-\frac{1}{\beta}\int_{\kp}{\sum_{n}}G_s(\kp,i\omega_n)G_s(\kp+\qp,i\omega_n+i\nu_m)\;.
\label{eq:chi-def}
\end{equation}
where $\int_{\kp}\!=\!\int dk_xdk_y/(2\pi)^2$, $\beta\!=\!1/(k_{\textrm{B}}T)$ is the inverse temperature, and 
($\mathbf{k}$,\,$\omega$) and ($\mathbf{q}$,\,$\nu$) denote Fermionic and Bosonic wavevectors and 
frequencies respectively. $\mathbf{k}_{\parallel}$ ($\mathbf{q}_{\parallel}$) represents an in-plane wavevector.
Plugging Eq.~(\ref{eq:gsf-spec}) in to Eq.~(\ref{eq:chi-def}), and summing over the Matsubara frequencies
using the standard contour technique, then taking the analytic continuation $i\nu_m\!\to\!\nu+i\delta$, 
one obtains
\begin{equation}
\Im\,\chi(\qp,\nu,\mu)\!=\!\int_{\kp}\int_{-\tmu}^{-\tmu+\tnu}\frac{d\epsilon}{t_2}\,f(-\epsilon,\kp)f(\tnu-\epsilon,\kp+\qp) 
\label{eq:chi-spec}
\end{equation}
where $\tmu\!=\!\mu/t_2$, and $\tnu\!=\!\nu/t_2$, with $\mu$ being the Fermi level. 
Since $f\!=\!f_s+f_b$, $\chi(\qp,\nu)$ can be decomposed into 
four terms which are the bulk-bulk ($\chi_{\textrm{bb}}$),
surface-bulk ($\chi_{\textrm{sb}}$), bulk-surface ($\chi_{\textrm{bs}}$) and surface-surface ($\chi_{\textrm{ss}}$) 
contributions. We will discuss these contributions separately in the following paragraphs.

\subsection{Partially filled surface bands}
\label{sec:part-fill-surf}

Let us first consider the situation with partially filled surface bands as shown in Fig.~\ref{fig:eh}(a) with $\mu\!<\!0$. 
The dynamical susceptibility contributed by the $s\!-\!s$ process (denoted by $\chi_{ss}(\qp,\nu)$) 
behaves similarly to the 2D Linhard function because the SGF has a pole at $\tw\!=\!0$ for $\kps\!<\!k_0$, 
which looks similar to that of 2D free electrons with quadratic dispersion. Thus the imaginary part of 
zero-temperature susceptibility $\textrm{Im}{\chi_{ss}(\qp,\nu)}\!\sim\!\nu/q_{\parallel}$ at small in-plane wavevector $q_{\parallel}$ and low frequency $\nu\ll \hbar v_{\textrm{F}}^{s}q_{\parallel}$ with $v_{\textrm{F}}^{s}$ referring to the Fermi velocity of the surface bands (In the finite-temperature formalism 
$\chi_{\textrm{ss}}(\qp, \nu_m)\!\sim\!\vert\nu_m\vert/q_{\parallel}$  with 
$\nu_m$ being Bosonic Matsubara frequency.). On the other hand, the dynamical susceptibility contributed by the $b\!-\!b$ process $\chi_{\textrm{bb}}(\mathbf{q}_{\parallel},\nu)$  with $\vert \nu\vert\!<\!\vert \mu\vert $  is expressed as:
\begin{equation}
\Im\,\chi_{\textrm{bb}}(\qp,\nu,\mu)\!=\!\int_{\kp}\int_{-\tmu}^{-\tmu+\tnu}\frac{d\epsilon}{t_2}\,f_b(-\epsilon,\kp)f_b(\nu-\epsilon,\kp+\qp)\;,
\label{eq:chibb}
\end{equation}

After some algebra, it turns out  that when $\nu\ll \hbar v_{\textrm{F}} q_{\parallel}$ ($v_{\textrm{F}}$ is the 
bulk Fermi velocity): 
\begin{equation}
\textrm{Im}\,\chi_{\textrm{bb}}(\qp,\nu,\mu)\sim \frac{\nu}{q_{\parallel}}\,.
\label{eq:chibbim}
\end{equation}
Therefore $\chi_{\textrm{bb}}$ is equally important as $\chi_{\textrm{ss}}$ for the hole-doped case.
In other words, the dominant Landau damping is from both the surface and the bulk, and they make comparable contributions.
Thus we expect the usual theory of 2d FM quantum critical still applies, with
consequently dynamical critical exponent $z \approx 3$.  It is also interesting to note that as a result of the fluctuations in the third spatial dimension, $\chi_{bb}(\qp,{\nu})$ is novanishing even when $q_{\parallel}\!=\!0$. It turns out that
\begin{equation}
\textrm{Im} \chi_{bb}(\qps\!=\!0,{\nu},\mu)\sim\nu\;, 
\label{eq:chibb0}
\end{equation}
which is unusual for a ferromagnetic phase transition. 
We refer the readers to Appendix \ref{appen:chianalytic} for the derivations 
of Eq.~(\ref{eq:chibbim}) and  Eq.~(\ref{eq:chibb0}).

\begin{figure}
\includegraphics[width=3.5in]{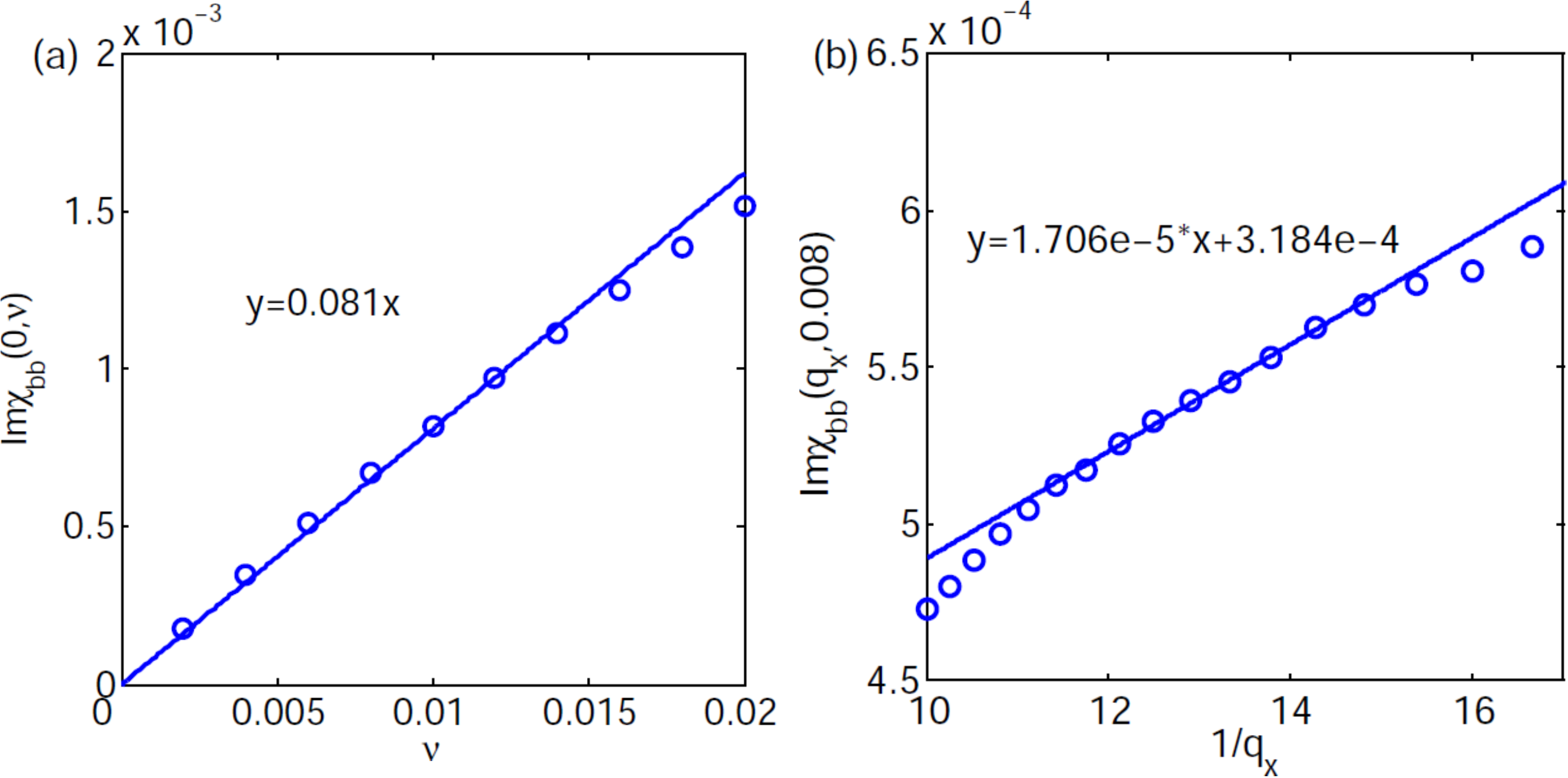}
\caption{Numerical calculations of the surface dynamical susceptibility of slightly hole-doped 
nodal-loop semimetals with partially filled surface bands: (a) frequency dependence at $\qps\!=\!0$; and 
(b) wavevector dependence at $\nu\!=\!0.008$. Note the horizontal axis in (b) is $1/q_x$.}
\label{fig:chibbnum}
\end{figure}

The analytic results shown in Eq.~(\ref{eq:chibb}) and Eq.~(\ref{eq:chibb0}) are supported by direct 
numeric calculations of the surface dynamical susceptibility  of a 500-cell slab of the tight-binding model
introduced in Sec.~\ref{sec:tbmodel}. 
The Fermi level $\mu\!=\!-0.036$ as schematically indicated by the gray dashed line in Fig.~\ref{fig:eh}(a), 
$t_0\!=\!0.01$, $t_1\!=\!0.8$, $t_2\!=\!0.3$ and $t_3\!=\!0.2$. The frequency dependence of surface dynamical susceptibility at 
$\qps\!=0\!$ is shown in Fig.~\ref{fig:chibbnum}(a). Clearly at low frequencies, $\chi_{\textrm{bb}}(0,\nu)$ is linear in $\nu$, in agreement with Eq.~(\ref{eq:chibb0}). 

We also study the wavevector dependence of $\chi_{\textrm{bb}}(\qp,\nu)$ for a given frequency $\nu\!=\!0.008$ as shown in Fig.~\ref{fig:chibbnum}(b).  
$\Im\,\chi_{\textrm{bb}}(\qp,\nu)$ is linearly dependent on $1/q_x$ for $0.065\lesssim q_{\parallel}\lesssim 0.085$ (in units of $1/a$, where $a\!=\!1$ is the in-plane lattice constant). When $\qps\lesssim 0.06$, we are no longer in the regime that $\nu\ll \hbar v_{\textrm{F}}q_{\parallel}$ and in the meanwhile $1/\qps$ becomes comparable to the $\k$-mesh density,   so that Eq.~(\ref{eq:chibbim}) is no longer valid; while when $q_{\parallel}$
is large ($q_{\parallel}\gtrsim 0.085$), the wavevector becomes comparable to the radius of the bulk ``Dirac cone" above which the electron-hole excitations are rigorously truncated. This explains why the 
$1/q_{\parallel}$ behavior is observed only for $0.065\lesssim q_{\parallel}\lesssim 0.086$.
The details of computing the surface dynamical susceptibility is explained in Appendix \ref{appen:chinum}.

\subsection{Nearly full surface bands}
\label{sec:nearly-full-surface}

We continue studying the case when the surface bands are nearly completely filled as shown in Fig.~\ref{fig:eh}(b). 
In such a situation, the Fermi level $\mu\!=\!0$, and the dominating contribution is either $b\!-\!b$ or $s\!-\!b$ process. 
The surface dynamical susceptibility from the $s\!-\!b$ process is expressed as
\begin{equation}
\textrm{Im}\,\chi_{\textrm{sb}}(\qp,\nu,\mu\!=\!0)=\int_{\kp}\int_{0}^{\tnu}\frac{d\epsilon}{t_2}\,
f_s(-\epsilon,\kp)f_b(\nu-\epsilon,\kp+\qp)\;,
\label{eq:chisb}
\end{equation}
and the $b-b$ contribution is expressed in Eq.~(\ref{eq:chibb}) with $\mu\!=\!0$.
After solving these integrals, 
it turns out that
\begin{align}
&\textrm{Im}\,\chi_{\textrm{bb}}(\qp, \nu,\mu\!=\!0)\sim \nu^3/\qps\;,
\label{eq:chibb3}\\
&\textrm{Im}\,\chi_{\textrm{sb}}(\qp, \nu,\mu\!=\!0)\sim \qps(\nu-\eta(t_0,\qps))\;,
\label{eq:chisb}
\end{align}
where $\eta(t_0,\qps)\!=\!2t_0(2k_0\qps-\qps^2)/3$ is the energy gap of the $s\!-\!b$
particle-hole excitations.
Physically Eq.~(\ref{eq:chisb}) implies that a minimal frequency $\sim\!\eta(t_0,\qps)$ 
is required to create an electron-hole pair of the $s\!-\!b$ type  with finite wavevector $\qps$. 
Such a minimal excitation energy $\sim\!t_0$, and vanishes when 
the surface bands are perfectly flat (remember that the surface bandwith arises due to $t_0$) or 
when $\qps\to\!0$. We refer the readers to Appendix.~\ref{appen:chianalytic} for the derivations
of Eq.~(\ref{eq:chibb3})-(\ref{eq:chisb}).
\begin{table}
\setlength{\tabcolsep}{3pt}
\caption{Linear fits to the frequency dependence of surface susceptibility at different wavevectors}
\begin{ruledtabular}
\begin{tabular}{lclclclclclclclclc}
$\qps$ & 0.4  & 0.3 & 0.25 & 0.2 & 0.15 & 0.1 & 0.05 \\
$c$   & 0.6075  & 0.5099  & 0.4536  & 0.3892  & 0.3143  & 0.2294 & 0.1377 \\
$\eta(t_0,\qps)$  & 0.0054   & 0.0041  & 0.0036  & 0.0030  & 0.0025  & 0.0017 & 0.0014
\end{tabular}
\end{ruledtabular}
\label{table:chisb}
\end{table}
Eq.~(\ref{eq:chibb3})-(\ref{eq:chisb}) indicate that when $\mu\!=\!0$ the $s\!-\!b$
process dominates over the $b\!-\!b$ process at low frequencies and
small wavevectors.   If we follow the Hertz-Millis procedure, a
straightforward analysis then predicts the
dynamical critical exponent $z \approx 1$.  Subtleties similar to
those in the purely 2D case may still occur here, of course, but this
result is sufficient to show that the quantum critical behavior at
this transition is fundamentally different from that of a purely 2D
itinerant ferromagnet.  We once again note that, when $\qps\!=\!0$,
$\textrm{Im}\,\chi_{bb}(0, \nu)$ is non-vanishing and
$\sim\!\nu^2$ for $\mu\!=\!0$ due to the Fermionic fluctuations in the $z$ direction.

\begin{figure}
\includegraphics[width=3.5in]{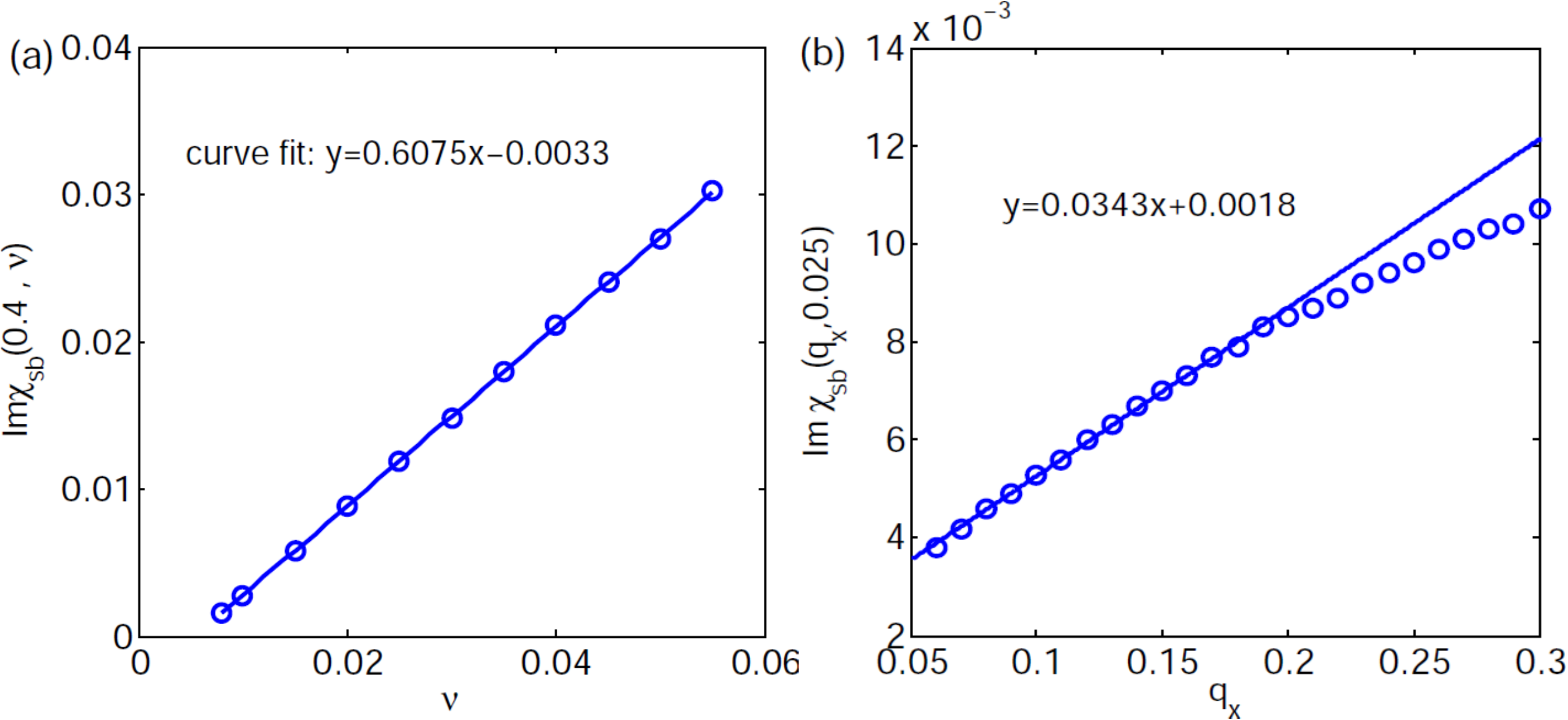}
\caption{Numerical calculations of the surface dynamical susceptibility of charge neutral
nodal-loop semimetals with nearly completely filled surface bands: (a) the frequency dependence at 
$\qps\!=\!0.4$, and (b) the wavevector dependence at $\nu\!=\!0.025$.}
\label{fig:chisbnum}
\end{figure}

Again, the analytic results in Eq.~(\ref{eq:chibb3})-(\ref{eq:chisb}) are numerically 
verified by directly computing the surface-layer dynamical susceptibility of a 500-cell slab.  
The Fermi level is very close to the nodal loop in the 
calculations as indicated by the gray dashed line in Fig.~\ref{fig:eh}(b). The surface bound states are almost completely
filled.  The other parameters of the tight-binding model are the same as those in the previous susceptibility calculation.  The frequency dependence of the surface susceptibility at $\qps\!=\!0.4$ (denoted by $\textrm{Im}\,\chi_{\textrm{sb}}(0.4, \nu)$) is shown in Fig.~\ref{fig:chisbnum}(a). Clearly 
$\textrm{Im}\,\chi_{\textrm{sb}}(0.4, \nu)\!\sim\!\nu$ at low frequencies and there is a small energy gap around $\nu\!\sim\!t_0$, 
in agreement with  Eq.~(\ref{eq:chisb}). 

In order to study the wavevector dependence of the energy gap $\eta(t_0,\qps)$, 
we have calculated the frequency dependence of the surface dynamical susceptibility of a 500-cell slab 
for different wavevectors from $\qps\!=\!0.4$ to $\qps\!=\!0.05$. Then we fit the data with linear  functions
$y\!=\!c\,(x-\eta(t_0,\qps))$ ($y$ is $\Im\,\chi_{\textrm{sb}}(\qps,\nu)$, $x$ is $\nu$).  
The parameters $c$s and $\eta(t_0,\qps)$s 
are shown  in Table \ref{table:chisb}. As clearly shown in the table, 
$\eta(t_0,\qps)$ decreases with $\qps$ and tend to vanish as $\qps\!\to\!0$
\footnote{Our simulations stop at $\qps\!=\!0.05$ because when $\qps\!\lesssim\!0.05$ the in-plane $\k$ mesh density
becomes comparable with $1/\qps$.}.

We also numerically calculate  the wavevector dependence of the 
surface dynamical susceptibility at $\nu\!=\!0.025$ as shown in Fig.~\ref{fig:chisbnum}(b). 
Clearly $\Im\,\chi_{\textrm{sb}}(\qps,0.025)\!\sim\!\qps$ at small $\qps$, 
in agreement with the analytic prediction of Eq.~(\ref{eq:chisb}). 
It should be noted that when the Fermi level is at the nodal energy, the $b\!-\!b$ process is suppressed at relatively large wavevector ($\qps\gtrsim\!0.05$), thus the data shown in Fig.~\ref{fig:chisbnum}(a)-(b) is mostly contributed by $s\!-\!b$ process.
We refer the readers to Appendix \ref{appen:chinum} for details in the the numeric calculations of surface dynamical susceptibility.

%

%%------------------------------------------------------
\section{Bulk quantum oscillations}
\label{sec:qo}
%%------------------------------------------------------

We turn to discussing  the bulk quantum oscillations of NLSMs neglecting Coulomb interactions.
%We consider a simplified situation that a circular nodal loop lies in the $k_x$-$k_y$ plane, and
%a magnetic field $\mathbf{B}\!=\!B\hat{e}_z$ normal to the nodal loop is applied.
We introduce the following low-energy effective
Hamiltonians describing nodal loops with different in-plane dispersions:
\begin{align}
&H_0^{\textrm{qua}}=\hbar v_z k_z\,\sigma_y+(\Delta-\frac{\hbar^2(k_x^2+k_y^2)}{2m})\,\sigma_z\;\nn
&H_0^{\textrm{lin}}=\hbar v_z k_z\,\sigma_y+(\Delta-\hbar v_0\sqrt{k_x^2+k_y^2})\,\sigma_z\;,
\label{eq:heff}
\end{align}
where $\sigma_y$ and $\sigma_z$ are the Pauli matrices representing the
lowest conduction band and highest valence band at some high-symmetry point ($\k=(0,0,0)$),
$v_z$ is the Fermi velocity along the $z$ direction, and $\Delta$ is the gap at $\k=(0,0,0)$.
$H_0^{\textrm{qua}}$ describes a circular nodal loop with quadratic in-plane dispersion, of which the in-plane effective
mass is denoted by $m$; while $H_0^{\textrm{lin}}$ describes a nodal loop with linear in-plane dispersion
with in-plane Fermi velocity $v_0$.   The nodal energies described by Eq.~(\ref{eq:heff})
are exactly zeros.

The Landau levels for the above two effective Hamiltonians with $\mathbf{B}\!=\!B\hat{e}_z$ 
are readily obtained:
\begin{align}
&E_{\pm}^{\textrm{qua}}(n,k_z)=\pm\sqrt{(\Delta-\hbar\omega_c(n+1/2))^2+\hbar v^2k_z^2}\;\nn
&E_{\pm}^{\textrm{lin}}(n,k_z)=\pm\sqrt{(\Delta- \hbar\omega_c\sqrt{n+1/2})^2+\hbar v^2k_z^2}\;,
\label{eq:ll}
\end{align}
where the cyclotron frequency 
\begin{equation}
\omega_c=
\begin{cases} eB/m & \hbox{for quadratic in-plane dispersion}\\ 
   \sqrt{2eBv_0^2/\hbar} & \hbox{for linear in-plane dispersion}
\end{cases}
\label{eq:wc}
\end{equation}
for the case of linear dispersion. If the chemical potential is exactly at the nodal energy, i.e., $\mu\!=\!0$, 
in general the Landau level spectrum is gapped and the chemical potential
is in the middle of the gap. However,
the gap closes at $k_z\!=\!0$ whenever 
$\Delta\!=\!\hbar\omega_c (n+1/2)$ for quadratic in-plane dispersion, and
$\Delta\!=\!\hbar\omega_c\sqrt{(n+1/2)}$ for linear in-plane dispersion.
Note that the above gap-closure condition is nothing but the equality between
the area of the nodal loop $\mathcal{A}_\textrm{NL}$ and the area of 
the $n$th quantized magnetic orbit $\mathcal{A}_{\textrm{B}}(n)$, i.e.,
$\mathcal{A}_{\textrm{NL}}\!=\!\mathcal{A}_{\textrm{B}}(n)$,
where $\mathcal{A}_{\textrm{NL}}\!=\!\pi \Delta^2/(\hbar^2v_0^2)$ ( 
$\mathcal{A}_{\textrm{NL}}=2\pi m\Delta/(\hbar^2)$) for a nodal loop 
with linear (quadratic) in-plane dispersions,
and the area of $n$th magnetic orbit $\mathcal{A}_{\textrm{B}}(n)\!=\!2\pi eB(n+1/2)/\hbar$. 

In other words, the Landau levels become gapless whenever the nodal loop 
exactly overlaps with a quantized magnetic orbit. At the gapless point there expects to 
be a sharp change in the free energy because
a fully occupied Landau level becomes completely unoccupied 
due to the gap closure and reopening. Thus some singular behavior is expected at the gapless 
critical point. 
%and one may interpret such a behavior as a magnetic-field-driven quantum phase transition
%
%\footnote{In Ref.~\onlinecite{nodal-chain-arxiv16}, it was  pointed out that the Berry phase of the 
%$n$th Landau level along the $k_z$ direction would be changed by $\pi$ through the gap closure 
%at $k_z=0$, thus such a quantum phase transition is indeed a topological phase transition.}.
%

To confirm the above conjecture, 
we calculate the magnetic susceptibility $\chi(B)\!=\!-\partial^2 F /\partial B^2$ for the Landau
levels shown in Eq.~(\ref{eq:ll}) in the limit $\mu\!\to\!0$ and $T\!\to\!0$.
It turns out that the magnetic susceptibility 
 consists a term which diverge logarithmically when  
the Landau level is gapless:
\begin{align}
\lim_{\mu\to 0,T\to 0}\chi(B)
\sim &\frac{e^2\omega_c}{\pi^2\hbar m v_z}
\sum_{n=0}^{\infty}(n+1/2)^2 \times \;\nn
&\ln\Big(\frac{\sqrt{(\Delta/\omega_c-(n+1/2))^2+\Lambda^2}+\Lambda}
{|\Delta/\omega_c-(n+1/2)|}\Big )
\label{eq:logdiv1}
\end{align}
for quadratic-inplane dispersion, and
\begin{align}
\lim_{\mu\to 0,T\to 0}\chi(B)\sim
&\frac{eB}{2\pi^2\hbar}\frac{e^2v_0^4}{\omega_c^2}\frac{2\Delta}{\omega_c v_z}
\sum_{n=0}^{\infty}\sqrt{n+1/2}\times \;\nn
&\ln\Big(\frac{\sqrt{(\Delta/\omega_c-\sqrt{n+1/2})^2+\Lambda^2}+\Lambda}
{|\Delta/\omega_c-\sqrt{n+1/2}|}\Big )
\label{eq:logdiv2}
\end{align}
for linear in-plane dispersion, where $\Lambda=(\pi v_z)/(\omega_c a)$ is a cutoff parameter
with $a$ being the lattice constant on the order of 1\,\angstrom.  Such logarithmic 
divergence indicates a
magnetic-field-driven quantum phase transitions in NLSMs 
\footnote{In Ref.~\onlinecite{nodal-chain-arxiv16}, it was  pointed out that the Berry phase of the 
$n$th Landau level along the $k_z$ direction would be changed by $\pi$ through the gap closure 
at $k_z=0$, thus such a quantum phase transition is indeed a topological phase transition.}.
More detailed results about the dHvA quantum oscillations of NLSMs are presented in Supplementary
Material.

%%-------------------------------------------------------------------------------
\section{Conclusion} 
\label{sec:summ}
%%---------------------------------------------------------------------------------

To summarize, we have studied the effects of Hubbard interactions and
bulk quantum oscillations in NLSMs. Our HF calculations indicate that
Hubbard interactions tend to drive the system into surface-ordered
phases through quantum phase transitions at the surface.  In
particular, in the absence of surface Rashba SOC, the system becomes
ferromagnetic at the surface at small $U$, and enters into a surface
charge-ordered phase at slightly increased $U$ through a first-order
transition.  On the other hand, surface Rashba SOC splits the
otherwise two-fold degenerate drumhead surface states and endows them
with nontrivial spin textures, so that a surface canted FM phase
becomes stable for moderate $U$ values.  The quantum critical behavior
of the surface ferromagnetic transition is distinct from that in
conventional 2D or 3D metals.  This is due to novel Landau damping of
the 2D spin fluctuations into electron-hole excitations near the nodal
loop in the third dimension.  This ``mixed dimensionality'' of the
system is argued to result in a modified dynamical critical exponent,
with $z\approx 1$ at the level of a Hertz-Millis analysis, when the
Fermi level is close to the bulk nodal energy.  We have also studied
the bulk quantum oscillations of NLSMs in the noninteracting case, and
find that in the limit of zero temperature and zero chemical
potential, there is a logarithmic divergence in the magnetic
susceptibility whenever the nodal loop overlaps with a quantized
magnetic orbit.  Such a logarithmic divergence is accompanied by the
gap closure of the Landau levels, and is periodic in $1/B$.  The
predictions of interaction-driven surface ordering and novel bulk
quantum oscillations may stimulate future experimental and theoretical
studies of NLSMs.

\textit{Note added.} Recently we became aware of two related works by
H. K. Pal \textit{et al.} \cite{pal-arxiv16} and B. Roy \cite{roy-16-nodal}.  In the former, the
authors have thoroughly studied the quantum-oscillation behaviors of various
physical quantities in a model of two dimensional valence and
conduction bands that touch along a loop, and in this context explored
the temperature dependence of the quantum oscillations. In the latter, the author
has discussed effects of Coulomb interactions in the bulk of nodal-loop semimetals.

\acknowledgments

JL would like to thank Se Young Park for his education on the
linear-tetrahedron method.  This work was supported by the National
Science Foundation under grant NSF DMR1506119.

\appendix
%--------------------------------------------------------
\section{Self-consistent Hartree-Fock approximation}
\label{appen:hf}
%--------------------------------------------------------
In Sec.~\ref{sec:hubbard}, the self-consistent Hartree-Fock (HF) 
approximation is adopted to calculate the ground states of the interacting Hamiltonians,
i.e., 
\begin{align}
U\hat{n}_{i\uparrow}\hat{n}_{i\downarrow}\to U\langle \hat{n}_{i\uparrow} \rangle \hat{n}_{i\downarrow}
+ U \hat{n}_{i\uparrow}  \langle \hat{n}_{i\downarrow} \rangle  
-U\langle \hat{n}_{i\uparrow} \rangle  \langle \hat{n}_{i\downarrow} \rangle 
\end{align}
where $\hat{n}_{i\sigma}$ refers to the density operator of electrons with 
spin $\sigma$ ($\sigma=\uparrow,\downarrow$) 
at site $i$,  $\langle \hat{n}_{i\sigma}\rangle$ is the self-consistent mean field applied to
the electrons of spin $-\sigma$ at site $i$; $U$ denotes the amplitude of the Hubbard repulsion. 
The linear tetrahedron method \cite{lt_prb94} is implemented as an 
interpolation scheme so that the self-consistent calculations can be carried out with improved numeric
efficiency. 

Including SOC, the noncollinear HF is slightly 
more complicated then its collinear version:
\begin{align}
Un_{i\uparrow}n_{i\downarrow}\to\,
&U [c_{i\uparrow}^{\dagger},  c_{i\downarrow}^{\dagger}] 
(\langle n_l \rangle -\mathbf{m}_l\cdot\mathbf{s}_l) [c_{l\uparrow},  c_{l\downarrow}]^{T}\;\nn
&-U\langle n_{l\uparrow} \rangle  \langle n_{l\downarrow} \rangle 
+U\langle c_{l\uparrow}^{\dagger}c_{l\downarrow}\rangle
\langle c_{l\downarrow}^{\dagger}c_{l\uparrow}\rangle \;,
\end{align}
where $c_{l\uparrow(\downarrow)}^{\dagger}$ ($c_{l\uparrow(\downarrow)}$) represents the 
creation (annihilation) operator of electrons at site $i$ with $\uparrow$ ($\downarrow$)
denoting electrons' spins. 
$n_{l\sigma}=c_{l\sigma}^{\dagger}c_{l\sigma}$ ($\sigma\!=\!\uparrow,\downarrow$) 
is the number operator at site $i$ with spin $\sigma$, and 
$n_l=n_{l\uparrow}+n_{l\downarrow}$ is the total number operator. $\langle ... \rangle$ represents the
expectation value of some operator in the HF ground state. 
$\mathbf{s}_l\!=\![s_{l}^x, s_{l}^y,  s_{l}^z]$ are the 
Pauli matrices representing an electron's spin at site $l$, which
couples to the self-consistent vector field $\mathbf{m}_l\!=\![m_{l}^{x}, m_{l}^y,  m_{l}^{z}]$, where
\begin{align}
&m_l^x=\langle c_{l\uparrow}^{\dagger}c_{l\downarrow}+c_{l\downarrow}^{\dagger}c_{l\uparrow}\rangle\;\nn
&m_l^y=i\langle c_{l\downarrow}^{\dagger}c_{l\uparrow}-c_{l\uparrow}^{\dagger}c_{l\downarrow}\rangle\;\nn
&m_l^z=\langle c_{l\uparrow}^{\dagger}c_{l\uparrow}-c_{l\downarrow}^{\dagger}c_{l\downarrow}\rangle
\end{align}
%
%
%--------------------------------------------------------
\section{Generalized RPA susceptibility}
\label{appen:rpa}

The generalized susceptibility in the random phase approximation (RPA) $\chi^{\RPA}$ can be expressed as
\cite{rpa-spinel}
\begin{align}
\chi^{\RPA}
=(1-\chi^{(0)}\mathbb{U})^{-1}\chi^{(0)}
\end{align}
where $\chi^{(0)}$ and $\mathbb{U}$ are the matrices representing the 
bare susceptibility and the Coulomb interactions respectively. 
To be specific, the bare susceptibility can be calculated from the 
noninteracting Green's function,
\begin{align}
\chi^{(0)}_{\alpha\beta l, \alpha ' \beta' l'}(\mathbf{q},i\nu_n)
=&-k_{\textrm{B}}T \int \frac{dk^2}{(2\pi)^2} \sum_{i\omega_n}
G^{(0)}_{\alpha' l', \alpha l}(\mathbf{k}, i\omega_{n})\times\;\nn
&\,G^{(0)}_{\beta l, \beta' l'}(\mathbf{k}+\mathbf{q}, i\omega_{n}+i\nu_n)\;,
\end{align}
where the $\alpha$,$\alpha'$,$\beta$ and $\beta'$ are the spin indices, while $l$ and $l'$ label the 
lattice sites in the slab; $\mathbf{k}$ is the wavevector of the noninteracting Bloch functions, and 
the sum over Matsubara frequency $\omega_n$ can be taken analytically in the basis that diagonalizes the
noninteracting Hamiltonian at each $\mathbf{k}$. $k_\textrm{B}$ is the 
Boltzman constant and $T$ is the temperature; $k_\textrm{B}T$ is fixed as 1/100 in the 
RPA calculations in Sec.~\ref{sec:hubbard}.  Note that in the nonordered phase without spin-orbit coupling, 
all kinds of spin fluctuations are equivalent to each other, i.e., 
$\chi^{(0)}_{\uparrow\uparrow l, \uparrow \uparrow l'}\!=\!\chi^{(0)}_{\uparrow\downarrow l, \uparrow\downarrow l'}
\!=\!\chi^{(0)}_{\downarrow\uparrow l, \downarrow\uparrow l'}\!=\!\chi^{(0)}_{\downarrow\downarrow l, \downarrow\downarrow l'}$.
With SOC included, terms like $\chi^{(0)}_{\uparrow\uparrow l, \downarrow\downarrow l'}$ are also allowed, and spin fluctuations
become anisotropic. 

The interaction matrix
for Hubbard interactions is defined as:
\begin{align}
\mathbb{U}^{l,l'}_{\beta\alpha',\alpha\beta'}=-(U\delta_{l,l'}\delta_{\beta' \alpha'}
\delta_{\beta\alpha}\delta_{\alpha,-\alpha'}-U\delta_{l,l'}\delta_{\alpha\beta'}
\delta_{\beta\alpha '}\delta_{\alpha, -\beta})
\label{eq:hubbardu}
\end{align}
The over minus sign on the right-hand-side (RHS) of Eq.~(\ref{eq:hubbardu}) is from the minus sign in the
time-ordered exponential of the $S$ matrix \cite{coleman-book}. The first term on the RHS of  Eq.~(\ref{eq:hubbardu}) 
represents a direct Coulomb interaction, while the second term is the exchange interaction. 
Then the matrix element of the static RPA spin susceptibility ($\chi_{zz}^{\textrm{RPA}}(\mathbf{q})$) is expressed as
\begin{align}
\chi_{zz}^{\textrm{RPA}}(\mathbf{q})_{l,l'}=
&\chi^{\RPA}(\mathbf{q})_{\uparrow\uparrow l,\uparrow\uparrow l'}-
\chi^{\RPA}(\mathbf{q})_{\downarrow\downarrow l,\uparrow\uparrow l'}-\;\nn
&\chi^{\RPA}(\mathbf{q})_{\uparrow\uparrow l,\downarrow\downarrow l'}+
\chi^{\RPA}(\mathbf{q})_{\downarrow\downarrow l,\downarrow\downarrow l'}
\end{align}

The eigenvalues of the RPA spin susceptibility at $t_2\!=\!t_1$ and 
$U\!=\!0.25t_1$ ($t_1$ and $t_2$ are defined in Sec.~\ref{sec:tbmodel}) are shown in Fig.~\ref{fig:chi}(a).
As discussed in Sec.~\ref{sec:hubbard}, the surface modes are much stronger than the bulk modes, and tend to
diverge at $\overline{\Gamma}$ as $U$ approaches some critical value $U_c$ indicating a continuous quantum 
phase transition at the surface.

%The real-space distribution of the eigenvectors of the RPA surface spin susceptibility at $\overline{\Gamma}$
%(denoted by $V_{zz}^{\textrm{surf}}(\Gamma)$) is shown in Fig.~\ref{fig:chirpa}(b). Clearly the acoustic (denoted by 
%blue diamonds) and optical (denoted by red circles) surface modes 
%are strongly localized at the two surfaces of the slab, and decays exponentially into the bulk in an 
%oscillationary manner. This is indeed consistent with the real-space charge distribution
%in the surface charge-ordered phase as shown in the inset of Fig.~2(a) in the main text.
%The two surface modes are degenerate because the charge fluctuations at two surfaces of a thick slab 
%have negligible interactions, so that they can be either completely the same or completely opposite, corresponding
%to the acoustic and optical modes respectively. 
 
%--------------------------------------------------------
%%%%
\section{Surface Green's function}
\label{appen:gsf}
%%%

In this section we derive the surface Green's function of NLSMs using the method reported in Ref.~\onlinecite{kalkstein-71}.
To be specific, using the Dyson equation, the surface Green's function $G_{s}(\kp, \omega)$ can be expressed as:
\begin{equation}
G_{s}=G_0+G_0 V G_{s}\;,
\label{eq:sgf-def1}
\end{equation}
where  $G_{s}$ is the full surface Green's function with the corresponding Hamiltonian $H$, 
$V\!=\!H-H_0$ is the potential difference between a crystal with and without a surface, and
$G_0$ is the noninteracting bulk Green's function.
In the basis of the ``hybrid Wannier functions" \cite{maryam-prb14} which are extended in the $x\!-\!y$ plane and localized in the
$z$ direction, Eq.~(\ref{eq:sgf-def1}) can be written as:
\begin{equation}
G_{s}(\kp,\omega)=G_0(\kp,\omega; 0)+G_0(\kp,\omega ;1) V(-1,0) G_{s}(\kp,\omega)\;,
\label{eq:sgf-def2}
\end{equation}
where $G_0(\kp,\omega; l)$ ($l$ is an integer labelling the primitive cells in the $z$ direction) 
is the bulk Green's function defined in the hybrid Wannier function basis:
\begin{equation}
G_0(\kp,\omega; l)=\int \frac{dk_z}{2\pi} e^{ik_zl} G_0(\mathbf{k},\omega)\;,
\label{eq:sgf-g0}
\end{equation}
and the bulk Green's function $G_0(\mathbf{k},\omega)$ is:
\begin{equation}
G_0(\mathbf{k},\omega)=\frac{-\tw\,\mathbb{I}_{2\times2}-(\tt_1+t_2\cos{k_z})\,\tau_x-t_2\sin{k_z}\,\tau_y}
{\tt_1^2+t_2^2+2\tt_1t_2\cos{k_z}-t_2^2\,\tw^2}\; .
\label{eq:gfbulk}
\end{equation}
In the above equation $\mathbb{I}_{2\times2}$ is the $2\times2$ identity matrix, $\tau_x$, $\tau_y$ and 
$\tau_z$ are the Pauli matrices defined in the sublattice space. $\tt_1$ and $\tw$ are defined in 
Eq.~(\ref{eq:tw}) in Sec.~\ref{sec:qc}.%, and $\mu$ is the Fermi level. 
If the bulk tight-binding model introduced in Sec.~\ref{sec:tbmodel} 
is truncated at sublattice $A$ with an ideal surface termination, 
the surface perturbation potential $V(-1,0)$ can expressed as
\begin{equation}
V(-1,0)=\begin{pmatrix}
0 & 0\\
-t_2 & 0
\end{pmatrix}\; .
\label{eq:v}
\end{equation}

Plugging Eq.~(\ref{eq:sgf-g0})-(\ref{eq:v}) into Eq.~(\ref{eq:sgf-def2}), one obtains:
\begin{equation}
G_{s}(\kp,\omega)_{1,1}=\frac{G_0(\kp,\omega; 0)_{1,1}}{1+t_2G_0(\kp,\omega;1)_{1,2}}\;.
\label{eq:sgf-def3}
\end{equation}
where 
\begin{equation}
G_0(\kp,\omega; 1)_{1,2}=-\int_{k_z} e^{ik_z}\frac{\tt_1+t_2e^{-ik_z}}{\tt_1^2+t_2^2+2\tt_1t_2\cos{k_z}-t_2^2\,\tw^2}\;,
\label{eq:g012-a}
\end{equation}
and
\begin{equation}
G_0(\kp,\omega;0)_{1,1}=\int_{k_z}\frac{-t_2\,\tw}{\tt_1^2+t_2^2+2\tt_1 t_2\cos{k_z}-t_2^2\,\tw^2}\;, 
\end{equation}
where $\int_{k_z}\!=\!\int_{0}^{2\pi}dk_z/(2\pi)$.
Again, $\tw$ is defined in Eq.~(\ref{eq:tw}).
Defining $\eta\!=\!e^{ik_z}$, the integral over $k_z$ in Eq.~(\ref{eq:g012-a}) can be replaced by an contour integral
around a unit circle in the complex plane of $\eta$, and can be solved exactly:
\begin{equation}
G_0(\kp,\omega; 1)_{1,2}=-\frac{\tt_1\eta_{+}+t_2}{\tt_1 t_2(\sqrt{\gamma^2-4})},
\label{eq:g012-b}
\end{equation}
and 
\begin{equation}
G_0(\kp,\omega;0)_{1,1}=-\frac{t_2\,\tw}{\tt_1 t_2}\frac{1}{\sqrt{\gamma^2-4}}\;,
\label{eq:g011}
\end{equation}
where
\begin{equation} 
\gamma\!=\!(\tt_1^2+t_2^2-t_2^2\,\tw^2)/(\tt_1 t_2)\;, 
\label{eq:gamma-def}
\end{equation}
and 
\begin{equation}
\eta_{+}\!=\!(-\gamma+\sqrt{\gamma^2-4})/2\;.
\label{eq:eta-def}
\end{equation}
From Eq.~(\ref{eq:g012-b}) one may notice that $G_0(\kp,\omega; 1)_{1,2}$ is real only if $\gamma^2-4\!>\!0$,
which implies that $G_{s}(\kp,\omega)_{1,1}$ may have a pole on the real axis only when $\gamma^2-4\!>\!0$. It follows
that $\gamma^2-4\!=\!0$ defines the bulk spectral edge:when $\gamma^2-4\!<\!0$, $\omega$ is in the bulk continuum;
while when $\gamma^2-4\!>\!0$, $\omega$ is in the bulk gap and there may be bound-state solutions. Then it is straightforward 
to show that:
\begin{equation}
\begin{cases}
\textrm{if $-\vert\kps^2-k_0^2\vert\!<\!\tw\!<\!\vert\kps^2-k_0^2\vert$,} & \hbox{$\omega$ in the bulk gap,}\\
\textrm{if $\tw\!>\!\vert\kps^2-k_0^2\vert$ or $\tw\!<\!-\vert\kps^2-k_0^2\vert$,} & \hbox{$\omega$ in the bulk continuum},
\end{cases}
\end{equation}
where $k_0$ characterizing the size of the nodal loop is defined in Eq.~(\ref{eq:k0}), and $\tw$ is defined in Eq.~(\ref{eq:tw}).

Plugging Eq.~(\ref{eq:g012-b}) and Eq.~(\ref{eq:g011}) into Eq.~(\ref{eq:sgf-def3}), we obtain:
\begin{equation}
G_s(\kp,\omega)=\frac{-\,\tw}{\tt_1}\frac{1}{(\sqrt{\gamma^2-4}+\gamma)/2-t_2/\tt_1}\;.
\label{eq:gs-form1}
\end{equation}
%
%Plugging Eq.~(\ref{eq:gamma-def}) into the above equation, 
Plugging $\tt_1\!=\!t_2(1+\kps^2-k_0^2)$ into Eq.~(\ref{eq:gamma-def}), considering the low-energy excitations around the nodal loop
so that $\kps^2-k_0^2$ and $\tw$ are small,  one obtains the final expression of the surface Green's function shown in Eq.~(\ref{eq:sgf}) by dropping some terms higher order in $\kps^2-k_0^2$ and $\tw$.

When $\omega$ is in the bulk gap,  Eq.~(\ref{eq:sgf}) can be re-expressed as:
\begin{align}
G_s(\kps,\omega)&=\frac{-\tw}{t_2}\frac{1}{(\,\sqrt{(\kps^2-k_0^2)^2-\tw^2}+\kps^2-k_0^2\,)}\;\nn
&\approx\frac{-\tw}{t_2}\frac{1}{(\,\vert\kps^2-k_0^2\vert\,(1-\tw^2/(\kps^2-k_0^2)^2)+\kps^2-k_0^2\,)}
\label{eq:gsf-surf1}
\end{align}
From the above equation we see that for $\kps\!<\!k_0$,
\begin{equation}
G_s(\kps,\omega)\approx\frac{(k_0^2-\kps^2)}{t_2\,\tw}\;,
\end{equation}
corresponding to the drumhead surface states at $\tw\!=\!0$.

From the above analysis we see that when the surface is terminated at sublattice $A$
there are drumhead surface states with dispersion $t_0(\kps^2-k_0^2)$ inside the projected nodal loop. 
On the other hand, when the surface is terminated at sublattice $B$, 
the role of $\tt_1$ and $t_2$ is interchanged, so that there are drumhead surface states only when $\tt_1\!>\!t_2$, 
i.e., outside the projected nodal loop ($\kps\!>\!k_0$). 
This explains the termination-dependent surface states as shown in 
Fig.~\ref{fig:surfband}(a)-(b).

%%%%----------------------------------------------------
\section{Derivations of surface dynamical susceptibility}
\label{appen:chianalytic}
%%%%----------------------------------------------------

%%%----------------------------------------------------------------------
\subsection{Derivations of Eq.~(\ref{eq:chibbim}), Eq.~(\ref{eq:chibb0})}
%%%-----------------------------------------------------------------------

We first derive the low-energy, long-wavelength behavior of 
the surface dynamical susceptibility of a hole-doped NLSM  
contributed by the extended bulk states projected at the surface, which are expressed by 
Eq.~(\ref{eq:chibbim}) and Eq.~(\ref{eq:chibb0}) in Sec.~\ref{sec:qc}. Such contributions
are labelled as ``$b-b$" in Fig.~(\ref{fig:eh})(a).
In principle we need to calculate the imaginary part of $\chi_{bb}(\qp,\nu,\mu)$ which
is expressed in Eq.~(\ref{eq:chibb}).

Again, we consider the situation 
that the nodal loop is centered at $(\pi,\pi,\pi)$ whose size is small compared
to the BZ. Then we expand $\tt_1$ around $(\pi,\pi)$ up to quadratic order of $\kps$
as shown in Eq.~(\ref{eq:tt1}).
Since we are interested in Fermi-surface fluctuations from the bulk continuum, we neglect the
dispersion from $t_0$, so
the spectral density of the bulk continuum $f_b$ becomes
\begin{equation}
f_{b}(\kp,\epsilon)\approx\frac{\sqrt{\epsilon^2- x_{\kp}^2}}{\epsilon}\theta(\vert\epsilon\vert-\vert x_{\kp}\vert)\;.
\label{eq:fb}
\end{equation}
Without loss of generality, the Bosonic wavevector 
$\qp$ is chosen to point along the $x$ direction, $\qp\!=\!(\qps,0)$. 
Then we define 
\begin{equation}
x_{\kp}=\kps^2-k_0^2\;.
\label{eq:xk}
\end{equation}
We also define 
\begin{align}
&\tmu=\mu/t_2\;,\nn
&\tnu=\nu/t_2\;\nn
&\tt_0=t_0/t_2\;.
\end{align}

Plugging the expression of $f_b$  in Eq.~(\ref{eq:fb}) into Eq.~(\ref{eq:chibb}), one obtains:
\begin{widetext}
\begin{align}
\Im\,\chi(\qp,\nu,\mu)&=\int_{\kp}\int_{-\tmu}^{-\tmu+\tnu}\,d\epsilon
\,\frac{\sqrt{\epsilon^2-x_{\kp}^2}\sqrt{(\tnu-\epsilon)^2-x_{\kp+\qp}^2}}{-\epsilon(\tnu-\epsilon)}\,
\theta(\vert\epsilon\vert-\vert x_{\kp}\vert)\,\theta(\vert\nu-\epsilon\vert-\vert x_{\kp+\qp}\vert)\;\nn
&=\int_{-\tmu}^{-\tmu+\tnu}\,d\epsilon\int_{-\epsilon}^{\epsilon}dx
\int_{-(\epsilon-\tnu)/(2k_0\qps)}^{(\epsilon-\tnu)/(2k_0\qps)}dy
\frac{2k_0\qps\sqrt{\epsilon^2-x^2}\sqrt{(\tnu-\epsilon)^2/(4k_0^2\qps^2)-y^2}}{-\epsilon(\tnu-\epsilon)\sqrt{1-(\,y-(x+\qps^2)/(2k_0\qps)\,)^2}}\;\nn
&\approx\int_{-\tmu}^{-\tmu+\tnu}\,d\epsilon\int_{-\epsilon}^{\epsilon}dx
\int_{-(\epsilon-\tnu)/(2k_0\qps)}^{(\epsilon-\tnu)/(2k_0\qps)}dy
\frac{2k_0\qps\sqrt{\epsilon^2-x^2}\sqrt{(\tnu-\epsilon)^2/(4k_0^2\qps^2)-y^2}}{-\epsilon(\tnu-\epsilon)}\;\nn
&=\int_{-\tmu}^{-\tmu+\tnu}\,d\epsilon\int_{-\epsilon}^{\epsilon}dx 
\frac{\sqrt{\epsilon^2-x^2}(\epsilon-\tnu)}{2k_0\qps\epsilon}\;\nn
&=\int_{-\tmu}^{-\tmu+\tnu}\,d\epsilon\,\frac{\epsilon(\epsilon-\tnu)}{2k_0\qps}\int_{-1}^{1}dx'\sqrt{1-x'^2}\;\nn
&=\frac{\pi}{4k_0\qps}(\epsilon^3/3-\tnu\epsilon^2/2)\Big\vert_{-\tmu}^{-\tmu+\tnu}\;\nn
&=\frac{\pi}{4k_0\qps}(\tmu^2\tnu-\tnu^3/6)\;,
\label{eq:imchibb-long}
\end{align}
\end{widetext}
where the second line of the above equation follows due to the heaviside $\theta$ function, and $y\!=\!(x+\qps^2)/(2k_0\qps)+\cos{\phi}$, with $\phi$ being the angle between $\kp$ and $\qp$. We have made the
approximation that $\sqrt{1-(\,y-(x+\qps^2)/(2k_0\qps)\,)^2}\!\approx\!1$ when going from the second to the 
third line in Eq.~(\ref{eq:imchibb-long}). The fourth line of Eq.~(\ref{eq:imchibb-long}) follows by 
using the integral identity:
\begin{equation}
\int_{-b}^{b}dy\sqrt{b^2-y^2}=\frac{\pi}{2}b^2\;,
\label{eq:integral}
\end{equation}
where $b=(\epsilon-\tnu)/(2k_0\qps)$. Finally in the fifth line we define $\epsilon x'\!=\!x$, and it follows that
$\Im\,\chi_{bb}(\qp,\nu,\mu)\!\sim\!\nu/\qps$. Eq.~(\ref{eq:chibbim}) is proved.

As discussed in the main text, the surface susceptibility is nonvanishing even at $\qp\!=\!0$ due to the 
bulk fluctuations. As expressed in Eq.~(\ref{eq:chibb0}), 
$\Im\,\chi(\qps\!=\!0,\nu,\mu)\!\sim\!\nu$ for $\mu\!<\!0$.  Using some similar tricks as those in Eq.~(\ref{eq:imchibb-long}), 
it is straightforward to show that
when $\qps\!=\!0$, 
\begin{widetext}
\begin{align}
\Im\,\chi(\qps\!=\!0,\nu,\mu)&=2\pi \int_{-\tmu}^{-\tmu+\tnu}\,d\epsilon\int_{-1}^{1}\,dx'
\,(\epsilon-\tnu)\sqrt{1-x'^2}\sqrt{1-(1-\tnu/\epsilon)^2 x'^2}\;\nn
&\approx\frac{4}{3}\pi(\vert\tmu\vert\,\tnu-\tnu^2/2)\;,
\end{align}
\end{widetext}
where the integral over $x'$ is approximated by a constant $2/3$.
Such an approximation is valid as long as the frequency is much smaller than the Fermi level, i.e.,$\nu\!\ll\!\vert\mu\vert$.
Thus Eq.~(\ref{eq:chibb0}) is proved.

%%%-------------------------------------------------------------------------
\subsection{Derivations of Eq.~(\ref{eq:chibb3})-(\ref{eq:chisb})}
%%%-------------------------------------------------------------------------

Now we turn to the case of Fig.~\ref{fig:eh}(b), i.e., the surface bands are filled and the 
electron-hole excitations are mostly contributed by the $b\!-\!b$ and $s\!-\!b$ process. 

Let us first consider the $b\!-\!b$ process. Since we are interested in the bulk-state fluctuations,
we neglect the dispersions from $t_0$ in the bulk continuum spectral density, 
i.e., $\tw\!\approx\!\omega/t_2$, and Eq.~(\ref{eq:fb}) applies. 
One may still use Eq.~(\ref{eq:imchibb-long}), except that now the Fermi level is right 
at the nodal enenergy $\mu\!=\!0$. Then it immediately follows from Eq.~(\ref{eq:imchibb-long}) that
$\Im\,\chi(\qp,\nu,\mu\!=\!0)\!\sim\!\nu^3/\qps$, which proves Eq.~(\ref{eq:chibb3}).

Next we consider the process that an electron is created in the bulk conduction band and a hole is left in the otherwise
occupied surface bands as denoted by $s\!-\!b$  in Fig.~(\ref{fig:eh})(b). 
Let us consider a simplified case that $t_0\!=\!0$ so that the surface bands are perfectly flat and completely occupied.
Then,
\begin{align}
&f_{b}(\kp,\epsilon)\approx\frac{\sqrt{\epsilon^2- x_{\kps}^2}}{\epsilon}\,\theta\,(\vert\epsilon\vert-\vert x_{\kps}\vert)\;,\nn
&f_{s}(\kp,\epsilon)\approx\vert x_{\kps}\vert\,\delta(\epsilon)\,\theta(-x_{\kps})\;.
\label{eq:fbfs-approx}
\end{align}
Plugging the above equation into Eq.~(\ref{eq:chisb}), one obtains:
\begin{widetext}
\begin{align}
\Im\,\chi_{\textrm{sb}}(\qp,\nu,\mu\!=\!0)&=\int_{\kp}\int_{0}^{\tnu}d\epsilon
\,f_{s}(-\epsilon,\kp-\qp)\,f_b(\tnu-\epsilon,\kp)\;\nn
&=\int_{\kp}\int_0^{\tnu}d\epsilon\vert x_{\kp-\qp}\vert\,\delta(-\epsilon)\,\theta(-x_{\kp-\qp})
\,\frac{\sqrt{(\tnu-\epsilon)^2-x_{\kp}^2}}{\tnu-\epsilon}\theta(\vert\tnu-\epsilon\vert-\vert x_{\kp}\vert)\;\nn
&\approx\int_{-\tnu}^{\tnu}dx\int_{(x+\qps^2)/(2k_0\qps)}^{1}\frac{d\cos{\phi}}{\sqrt{1-\cos^2{\phi}}}(2\qps\cos{\phi}\sqrt{k_0^2+x}-x-\qps^2)
\frac{\sqrt{\tnu^2-x^2}}{\tnu}\;\nn
&=2\qps k_0\int_{-\tnu}^{\tnu}dx\,\frac{\sqrt{\tnu^2-x^2}}{\tnu}\sqrt{1-(x+\qps^2)^2/(4k_0^2\qps^2)}-
\int_{-\tnu}^{\tnu}\frac{\sqrt{\tnu^2-x^2}}{\tnu}(x+\qps^2)\int_{(x+\qps^2)/(2k_0\qps)}^{1}\frac{d\cos{\phi}}{\sqrt{1-\cos^2{\phi}}}\;\nn
&\approx 2\qps k_0\int_{-\tnu}^{\tnu}dx\,\frac{\sqrt{\tnu^2-x^2}}{\tnu}\sqrt{1-(x+\qps^2)^2/(4k_0^2\qps^2)}\;\nn
&\approx 2\qps k_0\int_{-\tnu}^{\tnu}dx\,\frac{\sqrt{\tnu^2-x^2}}{\tnu}\;\nn
&=\pi k_0\qps\tnu.
\label{eq:imchisb-flat}
\end{align}
\end{widetext}
In the above equation, $x\!\equiv\!x_{\kp}\!=\!\kps^2-k_0^2$, and we have made the approximation  
$x_{\kp-\qp}=(\vert\kp-\qp\vert)^2-k_0^2\!\approx\!x+\qps^2-2k_0\qps\cos{\phi}$. We have used the integral identity,
$\int dx (1/\sqrt{1-x})=-2\sqrt{1-x}$, when going from the third to the fourth line; and we have dropped
the second term on the right hand side of the fourth line because it is higher order $\sim\!\qps^2\tnu$ or $\sim\!\tnu^2\qps$. 
Finally we have made 
the approximation $\sqrt{(1-(x+\qps^2)^2/(4k_0^2\qps^2))}\!\approx\!1$ from the fifth to the six line.
We see that the final result presented in Eq.~(\ref{eq:imchisb-flat}) is consistent 
with Eq.~(\ref{eq:chisb}) in the main text when $t_0\!=\!0$. 
It follows that when the $\mu\!=\!0$, the $s-b$ process dominate over the $b-b$ process, and leads to a dynamical critical
exponent $z\!\approx\!1$.

Now we consider the case of nonvanishing $t_0$, i.e., the surface bands are not perfectly flat, but with a 
bandwidth $\sim\!t_0$. Plugging Eq.~(\ref{eq:fsfb2}) into Eq.~(\ref{eq:chisb}), then integrating over $\epsilon$,
one obtains:
\begin{widetext}
\begin{equation}
\Im\,\chi_{\textrm{sb}}(\qp,\nu,\mu\!=\!0)=\int_{\kp}\vert x_{\kp-\qp}\vert\,\theta(-x_{\kp-\qp})\,
\theta(\vert\tnu+\widetilde{t_0}x_{\kp-qp}\vert)\,\frac{\sqrt{(\tnu+\widetilde{t_0}x_{\kp-\qp})^2-x_{\kp}^2}}{\tnu+\widetilde{t_0}x_{\kp-\qp}}\;.
\label{eq:imchisb-t0}
\end{equation}
\end{widetext}
where $x_{\kp}$ is defined in Eq.~(\ref{eq:xk}). Let us define $x\!\equiv\!x_{\kp}$ and $y\!\equiv\!x_{\kp-\qp}$.
Since $x$ is around 0, we make the following approximation to $y$:
\begin{align}
y&=x_{\kp-\qp}\;\nn
&=x-2\sqrt{x^2+k_0^2}\,\qps\cos{\phi}+\qps^2\;\nn
&\approx x-2k_0\qps\cos{\phi}+\qps^2\;.
\label{eq:y-approx}
\end{align}
Plugging Eq.~(\ref{eq:y-approx}) into Eq.~(\ref{eq:imchisb-t0}), and imposing the constriants on the 
limits of integrations from the two Heaviside $\theta$ functions, one obtains
\begin{widetext}
\begin{align}
\Im\,\chi_{\textrm{sb}}(\qp,\nu,\mu\!=\!0)&=\int_{x-2k_0\qps+\qps^2}^{0}\frac{dy}{2k_0\qps}
\int_{-\vert \tnu+\tt_0y \vert}^{\vert \tnu+\tt_0y \vert}\sqrt{(\tnu+\tt_0y)^2-x^2}\,\frac{-y}{\tnu+\tt_0y}\;\nn
&\approx-\frac{\pi}{{4k_0\qps}}\int_{-2k_0\qps+\qps^2}^{0}dy\,y\,(\tnu+\tt_0y)\;,
\label{eq:imchisb-t0-2}
\end{align}
\end{widetext}
where the second line of the above equation follows due the following approximation on the 
limit of integration of $y$:
\begin{equation}
\int_{x-2k_0\qps+\qps^2}^{0}\to\int_{-2k_0\qps+\qps^2}^{0}\;,
\label{eq:int-approx}
\end{equation}
and we have used the integral identity
\begin{equation}
\int_{-\vert \tnu+\tt_0y \vert}^{\vert \tnu+\tt_0y \vert}dx \sqrt{(\tnu+\tt_0y)^2-x^2}
=\frac{\pi (\tnu+\tt_0y)^2}{2}\;.
\end{equation}
Now we need to discuss two different situations: $\tnu+\tt_0y\!>\!0$, and $\tnu+\tt_0y\!<\!0$.
If $\tnu+\tt_0 y\!>\!0$, it follows from Eq.~(\ref{eq:imchisb-t0-2}) that
\begin{equation}
\Im\,\chi_{\textrm{sb}}^{>}(\qp,\nu)=\frac{\pi\tnu^3}{24k_0\qps}\;.
\label{eq:imchisb-gt}
\end{equation}
If $\tnu+\tt_0 y\!<\!0$, it turns out
\begin{equation}
\Im\,\chi_{\textrm{sb}}^{<}(\qp,\nu)\approx-\frac{\pi\tnu^3}{24k_0\qps}+\frac{\pi k_0\qps}{2}(\tnu-\eta(t_0,\qps)))\;.
\label{eq:imchisb-lt}
\end{equation}
Combining the above two equations,
\begin{align}
\Im\,\chi_{\textrm{sb}}(\qp,\nu)&=\Im\,\chi_{\textrm{sb}}^{<}(\qp,\nu)+\Im\,\chi_{\textrm{sb}}^{>}(\qp,\nu)\;\nn
&=\frac{\pi k_0\qps}{2}(\tnu-\frac{2t_0}{3}(2k_0\qps-\qps^2))\;.
\label{eq:imchisb-gap}
\end{align}
Eq.~(\ref{eq:imchisb-gap}) has the same analytic behavior as Eq.~(\ref{eq:imchisb-flat}) when $t_0\!=\!0$, although
the coefficients differ by a factor of 2. We attribute such a difference in the coefficients 
to the approximation shown in Eq.~(\ref{eq:int-approx}), and we believe it is not important because it dose not change
the analytic behavior of $\chi_{\textrm{sb}}$.
It is also clearly seen from Eq.~(\ref{eq:imchisb-gap}) that the 
excitation gap $\eta(t_0,\qps)\!=\!2t_0(2k_0\qps-\qps^2)/3$, which is proportional to $t_0$ and vanishes as
$\qps\!\to\!0$. This is also in agreement with our numeric simulations as shown in Table.~\ref{table:chisb}.

%%%%
\section{Numeric calculations of surface dynamical susceptibility in slab geometry}
\label{appen:chinum}
%%%%

In this section we explain the technical details in the numerical calculations of the surface dynamical susceptibility
for a slab of NLSMs, as shown in Fig.~\ref{fig:chibbnum} and Fig.~\ref{fig:chisbnum}.  
When both the surface Rashba SOC and Coulomb interactions are neglected, 
the system can be considered as spinless, and we use $l, l'$ to label the 
lattice sites in the $z$ direction in a slab of NLSMs.
The matrix element of zero-temperature dynamical susceptibility is expressed as:
\begin{equation}
\chi_{ll'}(\qp,\nu)=i\int \frac{dk_x dk_y}{(2\pi)^2}\int \frac{d\omega}{2\pi}
G^{(0)}_{l'l}(\kp,\omega)G^{(0)}_{ll'}(\kp+\qp,\omega+\nu)\;,
\label{eq:chinum}
\end{equation}
where the $G^{(0)}(\kp,\omega)$ is  the 
noninteracting Green's function for a slab of NLSMs which can expressed in matrix form as follows:
\begin{equation}
G^{(0)}(\kp,\omega)=V(\kp)G^{(0)}_{\textrm{diag}}(\kp,\omega)V^{\dagger}(\kp)
\end{equation}
where $G^{(0)}_{\textrm{diag}}$ is a $2N\!\times\!2N$ ($N$ is the number of primitive cells in the slab, and there 
are two sublattices in each primitive cell) diagonal matrix whose $j$th diagonal element
$G^{(0)}_{\textrm{diag}}(\kp,\omega)_{jj}=1/(\omega-\epsilon_j(\kp)+i\delta_{j,\kp})$, $\delta_{j,\kp}$ is 
an infinitesimal quantity which is greater than (less than) 0 if the eigenenergy $\epsilon_j(\kp)$ is occupied (unoccupied).
$V(\kp)$ is the eigenvector matrix of the Hamiltonian for the slab at $\kp$ (denoted by $H_{\textrm{slab}}(\kp)$):
$\sum_{l'}H_{\textrm{slab}}(\kp)_{l,l'}V_{l',j}(\kp)\!=\!\epsilon_{j}(\kp)V_{l,j}(\kp)$.
Then Eq.~(\ref{eq:chinum}) becomes
\begin{widetext}
\begin{equation}
\chi_{ll'}(\qp,\nu)=i\int \frac{dk_x dk_y}{(2\pi)^2}\int_{-\infty}^{\infty}\frac{d\omega}{2\pi}\sum_{j,j'=1}^{2N} 
\frac{W_{ll'jj'}(\kp,\qp)}
{(\omega-\epsilon_{j}(\kp)+i\delta_{j,\kp})(\omega +\nu-\epsilon_{j'}(\kp+\qp)+i\delta_{j',\kp+\qp})}\;,
\label{eq:chinum0}
\end{equation}
\end{widetext}
where the spectral weight $W_{ll'jj'}(\kp,\qp)$ is defined as
\begin{equation}
W_{ll'jj'}(\kp,\qp)=V_{l',j}(\kp)V_{l,j}^*(\kp)V_{l,j'}(\kp+\qp)V_{l',j'}^*(\kp+\qp)
\end{equation}
The integration over $\omega$ can be carried out by closing the contour 
in the upper half plane, then Eq.~(\ref{eq:chinum0}) becomes 
\begin{widetext}
\begin{equation}
\chi_{ll'}(\qp,\nu +i\delta)=\int \frac{dk_x dk_y}{(2\pi)^2}\sum_{j,j'=1}^{2N}\frac{W_{ll'jj'}(\kp,\qp)
\Big(\,\theta(\,\mu-\epsilon_j(\kp)\,)-\theta(\,\mu-\epsilon_{j'}(\kp+\qp)\,)\,\Big)}
{\epsilon_j'(\kp +\qp)-\epsilon_j(\kp)-\nu -i\delta}
\label{eq:chinum1}
\end{equation}
\end{widetext}
If the top-surface layer is labelled as the $0$th layer, then the surface susceptibility 
$\chi_{\textrm{surf}}(\qp,\nu)\!=\!\chi_{00}(\qp,\nu)$.
The numeric integrations over $k_x$, $k_y$ are replaced by discrete summations on a $280\!\times\!280$ $\k$ mesh, and the infinitesimal quantity $\delta$ is chosen as $0.001$ in our  numerical calculations. The number of primitive cells in the 
slab is $500$.  

%%%%%
\section{Bulk quantum oscillations}
\label{appen:qo}
%%%%%

In this section we derive the dHvA quantum oscillations of bulk NLSMs neglecting Coulomb interactions.
We consider two types of low-energy effective Hamiltonians of NLSMs as shown in Eq.~(\ref{eq:heff}).
The energies of $H_0^{\textrm{qua}}$ ($H_0^{\textrm{lin}}$) in Eq.~(\ref{eq:heff}) have quadratic (linear)
in-plane dispersions. The tight-binding model introduced in Sec.~\ref{sec:tbmodel} can be reduced to a $\k\cdot\mathbf{p}$
model around the center of the NLSM that is similar to $H_0^{\text{qua}}$; the terms  
linear in $\kp$ are killed by tetragonal symmetry. However, we would like to discuss both situations
($H_0^{\textrm{qua}}$ and $H_0^{\textrm{lin}}$)) for the sake of generality.

Landau levels are formed when a magnetic field is applied along the $z$ direction. The expressions of the Landau levels
for $H_0^{\textrm{qua}}$ and $H_0^{\textrm{lin}}$ are shown in Eq.~(\ref{eq:ll}). As discussed in Sec.~\ref{sec:qo},
the Landau levels become gapless whenever the nodal loop 
exactly overlaps with a quantized magnetic orbit. It is also mentioned that 
the gapless point there expects to be a sharp change in the free energy and the 
magnetic susceptibility show logarithmic divergence at zero temperature and zero Fermi level. 
In the remaining part of this section, 
we will explicitly derive the magnetic susceptibilities
$\chi(B)$ as expressed in Eq.~(\ref{eq:logdiv1})-(\ref{eq:logdiv2}).

The free energy of the Landau levels with chemical potential $\mu$ is expressed as:
\begin{equation}
F=-\frac{eB}{\beta 2\pi^2\hbar}\int_{-\pi}^{\pi} dk_z \sum_{n=0}^{\infty}\sum_{\lambda=\pm}
\log{( 1+e^{-(E_{\lambda}(n,k_z)-\mu)\beta} )}
\label{eq:f1}
\end{equation}
where the $\lambda\!=\!\pm$ label the branch of Landau levels, and the 
Landau levels $E_{\pm}(n,k_z)$ are expressed in Eq.~(\ref{eq:ll}) 
for both $H_0^{\textrm{qua}}$ and $H_0^{\textrm{lin}}$. Summing over $\lambda$, Eq.~(\ref{eq:f1})
becomes
\begin{equation}
F=-\frac{eB}{\beta 2\pi^2\hbar}\int_{-\pi}^{\pi} dk_z \sum_{n=0}^{\infty}
\log{g(E(n,k_z),\mu,\beta)}\;,
\label{eq:f2}
\end{equation}
where 
\begin{equation}
 g(E(n,k_z),\mu,\beta)=1+e^{-(E(n,k_z)-\mu)\beta}+e^{(E(n,k_z)+\mu)\beta}+e^{2\mu\beta}\;,
\label{eq:gfunc}
\end{equation}
$E(n,k_z)\!=\!E_{+}(n,k_z)$ (see Eq.~(\ref{eq:ll})), and $\beta\!=\!1/(k_{\textrm{B}}T)$.

Then it is straightforward to calculate the magnetic susceptibility $\chi(B)\!=\!-\partial^2 F/\partial B^2$:
\begin{align}
%\frac{\partial F}{\partial B}=&-\frac{e}{\beta 2\pi^2\hbar}\int_{-\pi}^{\pi} dk_z \sum_{n=0}^{\infty}
%\log{g(E(n,k_z),\mu,\beta)}\;\nn
%&-\frac{eB}{\beta 2\pi^2\hbar}\int_{-\pi}^{\pi} dk_z \sum_{n=0}^{\infty} 
%h(E(n,k_z),\mu,\beta)\frac{\partial E(n,k_z)}{\partial B}
\chi(B)=\frac{e}{2\pi^2\hbar}\int_{-\pi}^{\pi}dk_z\sum_{n=0}^{\infty}
(h_1+h_2+h_3)
\label{eq:pf2pb2}
\end{align}
where  
\begin{align}
&h_1=h(E(n,k_z),\mu,\beta)\,\frac{\partial E(n,k_z)}{\partial B}\;,\nn
&h_2=B\,h(E(n,k_z),\mu,\beta)\,\frac{\partial^2 E(n,k_z)}{\partial^2 B}\;,\nn
&h_3=B\,\frac{\partial h(E(n,k_z),\mu,\beta) }{\partial E(n,k_z)}\,(\frac{\partial E(n,k_z)}{\partial B})^2\;.
\end{align}
$h(E(n,k_z),\mu,\beta)$ is defined as follows
\begin{equation}
h(E(n,k_z),\mu,\beta)=\frac{e^{(E(n,k_z)+\mu)\beta}-e^{-(E(n,k_z)-\mu)\beta}}
{1+e^{(E(n,k_z)+\mu)\beta}+e^{-(E(n,k_z)-\mu)\beta}+e^{2\mu\beta} }\;.
\label{eq:hfunc}
\end{equation}

For NLSMs with quadratic in-plane dispersions, the Landau levels are defined in the first line of Eq.~(\ref{eq:ll}).
Then the partial derivatives of $E(n,k_z)$ with respect to $B$ are readily obtained:
\begin{align}
&\frac{\partial E(n,k_z)}{\partial B}=\frac{e\,(n+1/2)\,(\omega_c(n+1/2)-\Delta)}{m\sqrt{v^2k_z^2+(\Delta-\omega_c(n+1/2))^2}}\;,\nn
&\frac{\partial^2 E(n,k_z)}{\partial^2 B}=\frac{e\,(n+1/2)^2\,v^2 k_z^2}{m\,[v^2k_z^2+(\Delta-\omega_c(n+1/2))^2]^{3/2}}\;.
\label{eq:pepb}
\end{align}
Plugging Eq.~(\ref{eq:pepb}) into Eq.~(\ref{eq:pf2pb2}),one obtains that when $\mu\!=\!0$ 
and $\beta\!\to\!\infty$ ($T\!\to\!0$), one obtains the expression of the magnetic susceptibility:
%
%\begin{widetext}
\begin{align}
\chi(B)=
&\frac{2e^2}{2\pi^2\hbar m}\int_{-\pi}^{\pi} dk_z \sum_{n=0}^{\infty} 
\frac{(n+\frac{1}{2})\,(\omega_c(n+\frac{1}{2})-\Delta)}{E(n,k_z)}\;\nn
&\frac{e^2}{2\pi^2\hbar m}\int_{-\pi}^{\pi} dk_z \sum_{n=0}^{\infty} 
(n+\frac{1}{2})^2\,\frac{\omega_c}{E(n,k_z)}\;,\nn
&-\frac{e^2}{2\pi^2\hbar m}\int_{-\pi}^{\pi} dk_z \sum_{n=0}^{\infty}
(n+\frac{1}{2})^2\,\frac{\omega_c(\omega_c(n+\frac{1}{2})-\Delta)^2}{E(n,k_z)^3}
\label{eq:chiqua}
\end{align}
%\end{widetext}
%
The integration over $k_z$ in Eq.~(\ref{eq:chiqua}) can be carried out as follows:
\begin{align}
&\int_{-\pi}^{\pi} dk_z \frac{1}{E(n,k_z)}=\frac{2}{v}
\log\Big(\,\frac{\sqrt{((n+\frac{1}{2})-\frac{\Delta}{\omega_c})^2+\Lambda^2}+\Lambda}{\vert n+\frac{1}{2}-\frac{\Delta}{\omega_c}  \vert}\,\Big)\;\,\nn
&\int_{-\pi}^{\pi} dk_z \frac{1}{E(n,k_z)^3}=
\frac{2\Lambda}{v\omega_c^2(n+\frac{1}{2}-\frac{\Delta}{\omega_c})^2\sqrt{(n+\frac{1}{2}-\frac{\Delta}{\omega_c})^2+\Lambda^2}}
\label{eq:kzint}
\end{align}
where $\Lambda\!=\!\pi v/\omega_c$ is a dimensionless cutoff parameter (the in-plane lattice parameter is set to unity).

Plugging Eq.~(\ref{eq:kzint}) into Eq.~(\ref{eq:chiqua}), one obtains
\begin{align}
\chi(B)=&\frac{e^2\omega_c}{2\pi^2\hbar m}\sum_{n=0}^{\infty}\,\Big(\,(n+\frac{1}{2})^2\frac{2}{v}
\log(j(n,\omega_c,\Delta))\;\nn
&+2(n+\frac{1}{2})(n+\frac{1}{2}-\frac{\Delta}{\omega_c})\frac{2}{v}
\log(j(n,\omega_c,\Delta))\;\nn 
&-(n+\frac{1}{2})^2\frac{2\Lambda}{v\sqrt{(n+\frac{1}{2}-\frac{\Delta}{\omega_c})^2+\Lambda^2}}\,\Big)\;,
\label{eq:chiquafinal}
\end{align}
where 
\begin{equation}
j(n,\omega_c,\Delta)=
\frac{\sqrt{((n+\frac{1}{2})-\frac{\Delta}{\omega_c})^2+\Lambda^2}+\Lambda}{\vert n+\frac{1}{2}-\frac{\Delta}{\omega_c}  \vert}
\end{equation}
The first term on the RHS of Eq.~(\ref{eq:chiquafinal}) diverges logarithmically whenever $\Delta/\omega_c\!\to\!(n+1/2)$. On the other hand,
it is evidently seen that when $\Delta\!=\!(n+1/2)\omega_c$ is satisfied, the two Landau levels $\pm E(n,k_z)$ 
become gapless at $k_z\!=\!0$, and the size of the
quantized magnetic orbit associated with the $n$th Landau level becomes exactly the same as the size of the nodal loop.

One may reproduce the above derivations for a NLSM with linear in-plane dispersions (see $H^{\textrm{lin}}_0$ in Eq.~(\ref{eq:heff})). It turns out that for linear in-plane dispersions, the magnetic susceptibility is 
expressed as
\begin{align}
\chi(B)=
&\frac{eB}{2\pi^2\hbar}(\frac{ev_0^2}{\omega_c})^2\sum_{n=0}^{\infty}\sqrt{n+\frac{1}{2}}\frac{\Delta}{\omega_c}\frac{2}{v}
\log l(n,\omega_c,\Delta)\;\nn
&+\frac{e^2v_0^2}{\pi^2\hbar}\sum_{n=0}^{\infty}
\sqrt{n+\frac{1}{2}}(\sqrt{n+\frac{1}{2}}-\frac{\Delta}{\omega_c})\frac{2}{v}\log l(n,\omega_c,\Delta)\;\nn
&-\frac{eB}{2\pi^2\hbar}(\frac{ev_0^2}{\omega_c})^2\sum_{n=0}^{\infty}\frac{2\Lambda(n+\frac{1}{2})}
{v\omega_c^2\sqrt{(\frac{\Delta}{\omega_c}-\sqrt{n+\frac{1}{2}})^2+\Lambda^2}}\;,
\label{eq:chilin}
\end{align}
where 
\begin{equation}
l(n,\omega_c,\Delta)=\frac{\sqrt{(\frac{\Delta}{\omega_c}-\sqrt{n+\frac{1}{2}})^2+\Lambda^2}+\Lambda}
{\vert \frac{\Delta}{\omega_c}-\sqrt{n+\frac{1}{2}}\vert}\;.
\end{equation}
The first term on the RHS of Eq.~(\ref{eq:chilin}) diverge logarithmically whenever $\Delta\!=\!\omega_c\sqrt{n+1/2}$. 
Again, such a condition is exactly the gap-closure condition of Landau levels; in the meanwhile, the  $n$th magnetic
orbit exactly overlaps with the nodal loop when $\Delta\!=\!\omega_c\sqrt{n+1/2}$

\bibliography{nodal_loop}
\end{document}